\documentclass[a4paper,11pt]{article}
\pdfoutput=1 

\usepackage{jcappub}

\usepackage[T1]{fontenc} 

\usepackage{graphicx}
\usepackage{amsmath,amssymb}
\usepackage{braket}
\usepackage{subfigure}
\usepackage{float}
\usepackage[dvipsnames]{xcolor}
\usepackage{mathrsfs}
\usepackage{comment}
\usepackage{multirow}
\usepackage{soul}
\usepackage{mathtools}
\usepackage{sidecap} 
\graphicspath{{Figures/}}
\usepackage[normalem]{ulem}

\usepackage{tcolorbox}

\def\P{\mathcal{P}}

\newcommand{\be}{\begin{equation}}
\newcommand{\ee}{\end{equation}}
\newcommand{\llp}{\left [}
\newcommand{\rrp}{\right ]}
\newcommand{\lp}{\left (}
\newcommand{\rp}{\right )}
\newcommand{\bfx}{{\bf x}}
\newcommand{\bfy}{{\bf y}}

\newcommand{\bfp}{{\bf p}}

\newcommand{\bfk}{{\bf k}}
\newcommand{\bfq}{{\bf q}}
\newcommand{\bfu}{{\bf u}}

\newcommand{\dd}{{\mathrm d}}
\newcommand{\U}{\mathcal{U}}

\title{\boldmath Gravitational-wave lensing beyond rays: a disordered-system approach}

\author[a,b,1]{Ripalta Amoruso \note{Corresponding author},}
\author[c,d,1]{Ginevra Braga,}
\author[e]{Alice Garoffolo,}
\author[f,g,h]{Francescopaolo Lopez,}
\author[a,b,i]{Nicola Bartolo,}
\author[a,b,c,i]{Sabino Matarrese.}

\affiliation[a]{Dipartimento di Fisica e Astronomia "Galileo Galilei", Università degli Studi di Padova, via Marzolo 8, I-35131 Padova,Italy}
\affiliation[b]{INFN, Sezione di Padova,
via Marzolo 8, I-35131 Padova, Italy}
\affiliation[c]{Gran Sasso Science Institute, Viale F. Crispi 7, I-67100 L'Aquila, Italy}
\affiliation[d]{INFN-Laboratori Nazionali del Gran Sasso, Via G. Acitelli 22, 67100 Assergi (AQ), Italy}
\affiliation[e]{University of Pennsylvania
3451 Walnut St., Philadelphia, PA 19104, United States
}
\affiliation[f]{SISSA, International School for Advanced Studies, Via Bonomea 265, 34136 Trieste, Italy}
\affiliation[g]{INFN, Sezione di Trieste, Via Valerio 2, 34127 Trieste, Italy}
\affiliation[h]{IFPU, Istitute for Fundamental Physics of the Universe, Via Beirut 2, 34014 Trieste, Italy}
\affiliation[i]{INAF-Osservatorio Astronomico di Padova, Vicolo dell’ Osservatorio 5, I-35122 Padova, Italy}

\emailAdd{ripalta.amoruso@studenti.unipd.it}
\emailAdd{ginevra.braga@gssi.it}
\emailAdd{aligaro@sas.upenn.edu}
\emailAdd{flopez@sissa.it}
\emailAdd{nicola.bartolo@pd.infn.it}
\emailAdd{sabino.matarrese@pd.infn.it}

\abstract{We develop a framework to describe gravitational-wave propagation through a stochastic distribution of weak gravitational lenses beyond the geometric-optics limit. We model the lens distribution as a static random background field and formulate the problem in the language of quenched disorder, treating the disorder-averaged density matrix as the fundamental object from which observables are computed. Using the Schwinger–Keldysh formalism, we construct a path-integral representation of the averaged density matrix and derive its explicit form perturbatively for a suitable class of couplings. The result naturally separates into a quadratic exponential term, which governs the suppression of phase-sensitive contributions in the averaged description, and a purely oscillatory contribution, which modifies coherent propagation through a disorder-induced correction to the propagation kernel. This provides a unified description of interference, diffraction, and statistical fluctuations of the lens distribution within a single framework. We also identify the physical scales controlling the onset of coherence loss and illustrate the formalism in the case of Gaussian wave packets. More generally, the derivation applies to any system described by the same class of actions, making the framework relevant beyond gravitational-wave lensing to wave propagation in disordered media.}

\begin{document}
\maketitle

\section{Introduction}\label{sec:intro}

With more than two hundred gravitational-wave events observed by the LIGO–Virgo–KAGRA collaboration \cite{LIGOScientific:2018mvr,LIGOScientific:2020ibl,KAGRA:2021vkt,LIGOScientific:2025slb}, together with evidence for a stochastic gravitational-wave background reported by pulsar-timing-array collaborations \cite{NANOGrav:2023hde,EPTA:2023fyk,EPTA:2023sfo,EPTA:2023xxk,Xu:2023wog,Reardon:2023gzh,Reardon:2023zen}, gravitational waves (GWs) have emerged as a new probe of the Universe. Besides carrying information about their sources, they also probe the matter distribution they traverse. In this respect, propagation effects are not merely a nuisance, but a potential observational tool. Among them, gravitational lensing plays a central role, since it can leave observable distortions in the detected waveform, much as it does for electromagnetic signals.

Unlike photons, GWs span a broad range of wavelengths and may retain a high degree of coherence, depending on their source. Their interaction with intervening matter can therefore occur in qualitatively different regimes. When the wavelength is much smaller than the characteristic scale of the lens, the geometric-optics description applies: the wave is described in terms of rays, and lensing manifests itself through the focusing and defocusing of ray bundles~\cite{Isaacson:1968hbi,Isaacson:1968zza,Schneider:1992bmb}. When instead the wavelength becomes comparable to the lens scale, diffraction and interference become important. This is the wave-optics regime~\cite{Nakamura:1997sw,Nakamura:1999uwi}, where lensing corrections become frequency dependent, phase information becomes essential, and degeneracies between source and lens parameters can in principle be lifted~\cite{Cremonese:2021puh}. In particular, the event GW231123~\cite{LIGOScientific:2025rsn} has been identified as a possible candidate for lensing-induced wave-optics distortions and has already been analyzed within that framework~\cite{Goyal:2025eqo,Chan:2025kyu,Hu:2025lhv,Chakraborty:2025pxt}.

Most existing studies of GW lensing in wave optics have focused on single isolated lenses \cite{Feldbrugge:2019fjs,Braga:2024pik,Caliskan:2022hbu,Bonga:2024orc,Tambalo:2022wlm,Jow:2022pux,Villarrubia-Rojo:2024xcj,Savastano:2023spl,Tambalo:2022plm, Caliskan:2023zqm,Urrutia:2023mtk,Urrutia:2024pos,Urrutia:2021qak}, with only a few works considering a finite number of additional lenses~\cite{Yamamoto:2003cd,Feldbrugge:2020tti,Suvorov:2021uvd,Ramesh:2021nnl,Nakazono:2026vei}. The aim of this work is to move toward the more realistic situation in which the wave propagates through a random distribution of weak lenses. In this setting, the relevant observables are not associated with a single realization of the matter distribution, but with an ensemble of realizations, so averaging over the stochastic lens configurations becomes an essential part of the problem.

Such a statistical description is well established in the geometric-optics regime. There, the existence of well-defined trajectories makes it possible to treat random inhomogeneities either through line-of-sight approaches~\cite{Bertacca:2017vod,Laguna:2009re,Sasaki:1987ad,Hui:2005nm,Pyne:2003bn} or through kinetic descriptions based on Boltzmann equations~\cite{Contaldi:2016koz,Bartolo:2019yeu,LISACosmologyWorkingGroup:2022kbp,Bertacca:2019fnt,Bellomo:2021mer,Bravo:2025csu,Pitrou:2019rjz,Cusin:2017fwz}. In close analogy with CMB anisotropies, these methods can be used to study anisotropies of a stochastic GW background or to probe the statistical properties of matter with multiple resolved events. They have also been extended to theories beyond General Relativity and combined with electromagnetic-survey information to extract cosmological parameters~\cite{Flanagan:2008kz,Garoffolo:2019mna,Balaudo:2023klo,Garoffolo:2020vtd,Balaudo:2022znx,Tasinato:2021wol,Menadeo:2024uoq,Semenzato:2024mtn,Streibert:2024cuf,Ricciardone:2021kel,DeLeo:2025lmx,Mpetha:2022xqo,Yang:2023eqi}. In the wave-optics regime, however, no equally general framework is available. Some attempts in this direction can be found in~\cite{Pizzuti:2022nnj,Cusin:2018avf}, where wave effects are incorporated as a collision term in a Boltzmann equation. The basic obstacle is conceptual: once diffraction and interference become important, the notion of a ray is lost, and with it the standard trajectory-based techniques.

In this work, we address this problem by formulating GW propagation through a stochastic distribution of weak lenses in the language of disordered systems. More precisely, we model the lens distribution as a static random background, so that the problem naturally falls into the class of systems with {\it quenched disorder }~\cite{rammer,radzihovsky2015introduction,Cugliandolo2017AdvancedSP,Dotsenko:1995wv,bovier2006statistical,Ryzhik27R}. In such systems, the disorder does not introduce additional dynamical degrees of freedom, but acts as a fixed background through which the wave propagates. For any given realization, the evolution is unitary, while non-trivial statistical effects arise only after averaging over the ensemble of realizations. Quenched disorder plays a central role in statistical physics and condensed matter, where it provides the natural framework for wave propagation in media with frozen random inhomogeneities and underlies phenomena such as multiple scattering, weak localization, and disorder-induced coherence loss~\cite{vanRossum:1999zz,alma990047194370107871, Gneiting:2016pxb,Kropf:2016vto,Kropf:2017yzc,Gneiting:2020xrk}. It is also the setting in which Anderson localization was originally understood \cite{Anderson:1958vr}.

The natural object in this context is therefore the disorder-averaged density matrix. This is the quantity needed to compute expectation values of observables after averaging over the lens ensemble, while still retaining information about phase correlations and coherence. Intuitively, a statistical distribution of lenses induces random phase shifts during propagation. For a single realization, different paths interfere coherently and produce a definite interference pattern. Across different realizations, however, these phase relations fluctuate, so the corresponding interference patterns are no longer aligned. Ensemble averaging therefore suppresses phase-sensitive contributions, washes out interference, and reduces the off-diagonal coherence of the density matrix. In this sense, the stochastic lens distribution acts analogously to an environment: the evolution is unitary realization by realization, but becomes effectively non-unitary after disorder averaging.

Although the GWs considered here are classical, disorder techniques are natural because wave effects, which depend on amplitudes and phases, admit a close correspondence with quantum-mechanical language~\cite{2010EJPh...31..171M,Leung:2023lmq,Babington_crossing,goldstein2002classical,Schulman:1981vu}. We exploit this correspondence by treating the propagating GW as the system and the lenses as an external environment. Our strategy is to construct the density matrix for a fixed realization of the external potential, evolve it in time, and only afterward perform the ensemble average over disorder realizations. The resulting averaged density matrix encodes the effective interactions generated by the random medium and depends directly on the statistical properties of the lens distribution. This opens the possibility of probing quantities such as the correlation function or power spectrum of the lenses through their imprint on the observed GW signal, without assuming from the outset any particular optical regime.

More generally, the framework developed here is not restricted to GW lensing. It applies to any system in which wave degrees of freedom interact with a static random background through the class of couplings considered in this work. Since such interactions arise broadly in wave propagation through disordered media, the formalism may also be relevant for electromagnetic, ocean, and seismic waves propagating through random environments \cite{Eve:PI-wavetheory,Schlottmann:1999,1457032,843670}.

\bigskip
The paper is organized as follows. In Section~\ref{sec:Setup} we introduce the physical setup, describing the stochastic matter distribution, the effective wave equation for the propagating GW, and the motivation for adopting a density-matrix formulation. In Section~\ref{sec:Path Integral} we formulate the problem in the Schwinger–Keldysh path-integral language and define the disorder-averaged density matrix. In Section~\ref{sec:LensStat} we specify the statistical properties of the lens distribution and introduce the correlator and power spectrum of the gravitational potential. In Section~\ref{sec:The time evolution operator} we compute the time-evolution operator for a fixed realization of the disorder, separating the classical contribution from the fluctuation determinant and evaluating both perturbatively. In Section~\ref{sec:Density matrix results} we perform the disorder average and derive the explicit form of the averaged density matrix, discussing its physical interpretation in terms of dephasing, coherence loss, and effective modifications of the propagation kernel. Technical details are collected in the appendices.

\section{Setup}\label{sec:Setup}

\subsection{Background metric: the matter structures}
We consider a wave propagating in a Universe containing static matter inhomogeneities, modeled as a stochastic random field, treated as externally prescribed while neglecting any backreaction from the wave on the matter distribution. 
We therefore consider a Universe  described by the line element
\be \label{eq:BackgroundMetric}
\dd s^2 = -(1+ 2 \alpha \Phi(\bfx)) \dd t^2 + (1 - 2 \alpha \Phi(\bfx)) \delta_{ij} \dd x^i \dd x^j\,,
\ee
where $\Phi(\bfx)$ is the gravitational potential generated by the matter structures, assumed to be time independent for simplicity\footnote{In general cosmological settings the gravitational potential $\Phi$ is time dependent. In this work, however, we restrict attention to static realizations of the matter distribution. This provides a simplified but consistent framework in which the wave propagates through a fixed inhomogeneous background, and it is also the appropriate assumption for comparison with quenched disorder systems, where the environment is by definition static.}. We use the notation $x^\mu=(t_x,\bfx)$, with $\bfx$ denoting the spatial coordinates. 
We focus on GWs with sub-horizon wavelengths, so to neglect cosmological expansion. This way the metric reduces to the usual Newtonian form, appropriate for weak and static perturbations on sub-horizon scales. 
Within the regime of validity of the weak field approximation, the gravitational potential typically has amplitude $\Phi/c^2 \ll 1$. We will exploit the smallness of this quantity to develop a perturbative treatment, regarding the effect of matter inhomogeneities on the propagating wave as a weak correction to the unperturbed dynamics. To this end, in Eq.~\eqref{eq:BackgroundMetric}, we have introduced the bookkeeping parameter $\alpha$ to keep track of the perturbative order in the gravitational potential; it can be set to $1$ at the end of the calculation.

\subsection{The waves}
We are interested in the propagation of GWs on the background geometry given in Eq.~\eqref{eq:BackgroundMetric}. Treated in full generality this is a highly nontrivial problem, and a solution is in general only possible if suitable approximations are introduced. Since our main focus is on the low-frequency regime, the standard high-frequency approximation cannot be employed. That framework provides a systematic description of wave propagation on generic backgrounds, including polarization effects~\cite{Isaacson:1968hbi,Isaacson:1968zza}. As a first step, we therefore neglect polarization and model the GW as a real scalar field. Although this approximation is not fully realistic (see, for instance, \cite{Cusin:2019rmt,Braga:2024pik,Garoffolo:2022usx,Dalang:2021qhu} for analyses incorporating polarization effects in wave-optics and beyond geometric-optics) it allows us to isolate more clearly the basic mechanisms induced by the stochastic distribution of lenses.

Under this approximation, the GW is described by a real scalar field $\zeta$ obeying the sourceless wave equation on the curved background,
\be \label{eq:KGgeneral}
\frac{1}{\sqrt{-g}} \partial_\mu \lp \sqrt{-g} g^{\mu\nu } \partial_\nu \zeta \rp = 0\,,
\ee
where $g_{\mu\nu}$ is the metric in Eq.~\eqref{eq:BackgroundMetric}. Expanding to linear order in the weak gravitational potential, Eq.~\eqref{eq:KGgeneral} reduces to
\be
\label{eq:KG}
    \llp \partial_{\bfx}^2 - \lp 1 - 4 \alpha \Phi ( \bfx ) \rp \partial_t^2  \rrp \zeta(\bfx, t)=0\,.
\ee

Eq.~\eqref{eq:KG} provides the basic starting point for wave-optics analyses of gravitational lensing (see references in the Introduction), and it fully determines the final wave given the one realization of the gravitational potential.
Diffraction and interference are described by studying the solution of Eq.~\eqref{eq:KG} in terms of the so-called {\it diffraction integral}~\cite{Nakamura:1999uwi} which shows that these effects become important when the wavelength is comparable to, or larger than, the characteristic size of the lens (for instance for a point-mass lens this is the Schwarzschild radius). The diffraction integral is a path-integral solution of Eq.~\eqref{eq:KG}, valid under the assumptions of the {\it eikonal} and {\it paraxial} approximations, and for Newtonian lenses. Owing to these approximations, the path integral is expressed in terms of trajectories rather than field configurations, which makes the geometric-optics limit particularly transparent, since it emerges from the saddle-point approximation of the path integral. Given the observational prospect, considerable effort is being devoted to understanding how to go beyond its assumptions, such as: the worldline formalism~\cite{Gutzwiller_1967,Edwards:2019eby,Bastianelli:2006hq,Feynman:1950ir,Fishman1984,Guzman:2026jfr,Mogull:2020sak}, scattering-amplitude techniques~\cite{CarrilloGonzalez:2025gqm,Luna:2023uwd,Takahashi:2005sxa,DiVecchia:2023frv,Bjerrum-Bohr:2025bqg,Bautista:2021inx,Caron-Huot:2025tlq,Ivanov:2024sds,Nambu:2019sqn,Buoninfante:2024yth,Carrillo-Gonzalez:2021mqj, Nambu:2015aea}, black hole perturbation theory ~\cite{Pijnenburg:2024btj,Saketh:2025cwf,Chan:2025wgz}.

In this work, we aim to generalize the lensing description in wave optics considering many lenses. 
However, our scope is even broader.  As already mentioned in the Introduction, many other physical systems obey Eq.~\eqref{eq:KG}. For instance, in the context of optics, the gravitational potential $\Phi$ is replaced by the refractive index of the medium \cite{Eve:PI-wavetheory,Hannay:pathlinking,Fishman_2006,linares_2025_t6w3m-snb16}, while for acoustic or seismic waves, the role of $\Phi$ is played by the pressure field and density profiles \cite{Schlottmann:1999,1457032}.

\subsection{The need for a density-matrix description}
Our goal is not to describe propagation in a single realization of the lens distribution, but to obtain a statistical description of wave propagation through a static random medium. A conceivable strategy would be to solve Eq.~\eqref{eq:KG} for a fixed realization of the gravitational potential, construct from these solutions the relevant two-point quantity, and only afterward average over the stochastic ensemble of lenses. Although formally legitimate, this is not the most suitable approach for the present problem.

The reason is both technical and conceptual. Explicit solutions based on the diffraction integral are usually made tractable through the thin-lens approximation~\cite{Nakamura:1997sw,Nakamura:1999uwi}, which reduces the path integral to an effective two-dimensional integral on the lens plane. Here, however, we wish to describe an extended lens distribution without assuming that the interaction is localized on a single plane or that successive scattering events are independent. Once the thin-lens approximation is abandoned, the diffraction integral formalism is replaced by the full infinite-dimensional path integral, and it becomes more natural to formulate the problem directly at the level of the field configuration. The appropriate object is therefore the {\it averaged density matrix}, which encodes both occupation probabilities and phase correlations, and therefore directly captures interference and coherence. 

Before proceeding, let us note that related questions of fields interacting with environments have been widely studied in cosmology in the framework of {\it open quantum systems}, rather than quenched disorder~\cite{Breuer:2002pc,Martin:2018zbe,Boyanovsky:2015jen,Burgess:2024eng,Colas:2023wxa,Colas:2022kfu,Colas:2024xjy,Salcedo:2025ezu,Lopez:2025arw,Colas:2024ysu,Christie:2025knc,Zarei:2021dpb,Sharifian:2023jem,deKruijf:2024ufs,Takeda:2025cye,Burgess:2022nwu,Burgess:2025dwm}. The two frameworks are conceptually close, since both take the density matrix as the basic object, but they differ in the nature of the environment: non-dynamical in disordered systems, fully dynamical in open quantum systems. In both cases, however, the central issue is how the environment modifies the reduced dynamics of the system, and in particular its coherence properties.


\section{The averaged density matrix}

In this section we introduce the averaged density matrix and derive its path integral representation for a given initial GW configuration.  Although we will use a language typical of quantum field theory, our treatment is not meant to quantize the GW field in a fundamental sense. Rather, the quantum formalism provides a convenient way to describe the statistical propagation of a coherent classical wave through a random medium, keeping track of both amplitudes and phase correlations. For this reason, the quantity that appears in the phase of the path integral will not be identified with $\hbar$, but with a parameter $\Omega$ that plays the analogous role of a large semiclassical scale and will later be related, through the correspondence principle, to the wave frequency \cite{2010EJPh...31..171M}.

\subsection{Path-integral description of the averaged density matrix}\label{sec:Path Integral}

We begin by considering a fixed realization of the gravitational potential $\Phi(\bfx)$. For such a realization, the dynamics is described by the action
\be \label{eq:action in spacetime}
S[\zeta,\Phi]=S_0[\zeta]+S_{\rm int}[\zeta,\Phi] \,,
\ee
with
\begin{align}
S_0 &= \int_{t_i}^{t_f} \dd^4x \: \partial_\mu \zeta(\bfx,t)\partial^\mu \zeta(\bfx,t)\,, \\
S_{\rm int} &= 4\alpha \int_{t_i}^{t_f} \dd^4x \: \Phi(\bfx)\,\partial_t \zeta(\bfx,t)\partial_t \zeta(\bfx,t)\,,
\end{align}
which produces Eq.~\eqref{eq:KG} as equation of motion. In these formulas, indices are raised and lowered with the Minkowski metric $\eta_{\mu\nu}$. To describe propagation over a finite time interval, we restrict the time integration to $t\in[t_i,t_f]$, so that
\begin{equation}
\int_{t_i}^{t_f}\dd^4x=\int_{t_i}^{t_f}\dd t_x\int \dd^3\bfx \,.
\end{equation}
Here $S_0$ governs the free propagation of the wave, while $S_{\rm int}$ describes its coupling to the external gravitational potential.

For each fixed realization $\Phi$, the state of the system is described by a density operator $\hat\rho^\Phi(t)$. Since $\Phi$ is treated as a prescribed external field, the evolution for fixed realization is unitary. In the Schr\"odinger representation one therefore has
\be\label{eq:TimeEvolutionDM}
\hat \rho^\Phi(t)=\hat \U_\Phi(t,t_i)\,\hat \rho(t_i)\,\hat \U_\Phi^\dagger(t,t_i)\,,
\ee
where $\hat \U_\Phi(t,t_i)$ is the time evolution operator in the background $\Phi$, which is unitary. We assume that the initial density operator $\hat\rho(t_i)$ is independent of the realization of the gravitational potential. This is natural when the initial state is prepared independently of the lens configuration (as for instance for the cosmological GW background), while this assumption may need to be revisited in situations where source and lens environment are correlated (such as the  GW background generated by astrophysical sources).

The matrix elements of the evolution operator admit the path-integral representation
\be\label{Upath}
\U_{\Phi}(\zeta_f,\zeta_i; t_t, t_i)
=\langle \zeta_f|\hat \U_\Phi(t_f,t_i)|\zeta_i\rangle
= \int_{\zeta_i}^{\zeta_f}\mathcal{D}\zeta\, e^{i\Omega S[\zeta,\Phi]}\,,
\ee
where the initial and final field configurations $\zeta_i$ and $\zeta_f$ are defined by
\begin{align}\label{eq:BC}
\zeta(\bfx,t_i)= \zeta_i(\bfx)\,, \qquad
\zeta(\bfx,t_f)= \zeta_f(\bfx)\,,
\end{align}
and the action is evaluated with time integration bounded between $t_i$ and $t_f$. In Eq.~\eqref{Upath} we have introduced the dimensionful constant $\Omega$, taking the place of $\hbar^{-1}$, so that the exponent is dimensionless, as  anticipated.

Since the evolution of a density matrix involves both forward and backward time propagation, it is natural to employ the {\it Schwinger--Keldysh}, or in-in formalism~\cite{Schwinger:1960qe,Keldysh:1964ud}. In contrast to the usual in-out approach, the in-in formalism is designed to compute expectation values of operators at finite times and is therefore particularly suited to systems out of equilibrium.
In this representation, the time evolution is defined along a closed time contour that runs forward from the initial time to the final time and then backward to the initial time. As a consequence, the field content is doubled, with two copies of the field: $\zeta^+$ living on the forward branch and $\zeta^-$ on the backward branch.
The corresponding boundary data are $\{\zeta_i^+,\zeta_i^-\}$ at the initial time and $\{\zeta_f^+,\zeta_f^-\}$ at the final time. 
In the context of cosmology, this formalism was introduced and implemented in \cite{Calzetta:1986cq,Calzetta:1986ey,Calzetta:1995ys,Calzetta:2008iqa} (see also \cite{Weinberg:2005vy}).

For a fixed realization of the gravitational potential, the matrix elements of the evolved density operator are therefore
\begin{align}\label{Eq:density_matrix_evolved}
\rho^\Phi(\zeta_f^+,\zeta_f^-;t) &\equiv \langle \zeta_f^+|\hat \rho^\Phi(t)|\zeta_f^-\rangle  = \langle \zeta_f^+|\hat \U_\Phi(t,t_i)\,\hat \rho(t_i)\,\hat \U_\Phi^\dagger(t,t_i)|\zeta_f^-\rangle \,.
\end{align}
The quantity of interest is therefore not $\rho^\Phi$ for a single realization, but the disorder-averaged density matrix~\cite{Kropf:2016vto},
\begin{align}
\label{densitymatrixevolution}
\rho_{\rm av}(\zeta_f^+,\zeta_f^-;t) &= \int \mathcal D\Phi\,\P[\Phi] \int \dd \zeta_i^+\,\dd \zeta_i^- \, \U_\Phi(\zeta_f^+,\zeta_i^+;t,t_i)\, \rho(\zeta_i^+,\zeta_i^-;t_i)\, \U_\Phi^\dagger(\zeta_f^-,\zeta_i^-;t,t_i)\,,
\end{align}
where $\P[\Phi]$ is the probability density functional of the stochastic gravitational potential.  Note that, while the scalar field is doubled along the Schwinger--Keldysh contour,  the stochastic potential is not: the same realization $\Phi(\bfx)$ appears on both branches.

In the field basis, the diagonal elements $\rho^\Phi(\zeta,\zeta;t)$ encode the statistical weight of the different field configurations, while the off-diagonal elements $\rho^\Phi(\zeta,\zeta’;t)$ with $\zeta\neq\zeta’$ encode the phase correlations between them. Eq.~\eqref{densitymatrixevolution} shows that the disorder average couples the forward and backward evolutions, since the same realization of the gravitational potential enters both branches. This coupling is the imprint of the random medium on the effective dynamics and is responsible for the emergence of non-unitary features in the disorder-averaged description. In particular, it can suppress the off-diagonal elements of the density matrix, leading to a loss of coherence. In the present setup, however, this effect is not produced by a dynamical bath, as in open quantum systems, but by averaging over a static random background. For each fixed realization of the gravitational potential the evolution remains unitary, so no intrinsic loss of coherence occurs; the suppression of off-diagonal terms arises only after ensemble averaging. Dephasing is therefore not a property of propagation in a single background, but an emergent effect of quenched disorder.

\subsection{Observables}

Once the disorder-averaged density matrix has been determined, it can be used to compute the expectation value of any observable acting on the propagating wave. 
Indeed, let $\hat A$ be an operator on the Hilbert space of the field, then 
\be
 \langle \hat A(t_f)\rangle_{\rm av} = \int {\cal D}\Phi\,{\cal P}[\Phi]\,
{\rm Tr} \llp \hat\rho^\Phi(t_f)\,\hat A \rrp = {\rm Tr} \llp \rho_{\rm av} (t_f) \; \hat A \rrp \,,
\ee
since both the disorder average and the trace are linear operations. Thus, the averaged density matrix is used to time evolve the average operators, and finding their expectation values. The trace operation in this case is defined as 
\be 
{\rm Tr} \llp \hat \rho \, \hat A \rrp \equiv  \int \dd\zeta_f^+\,\dd\zeta_f^- \; \rho (\zeta_f^+,\zeta_f^-;t)\, A(\zeta_f^-,\zeta_f^+;t) \,, \qquad A(\zeta_f^-,\zeta_f^+;t)\equiv \langle \zeta_f^-|\hat A (t) |\zeta_f^+\rangle \,.
\ee 
Therefore, one may either first evolve the density matrix for each realization and then average, or first construct the averaged density matrix and then evaluate the observable. The two procedures are equivalent~\cite{Gneiting:2020xrk}.

If one is interested specifically in GW observable,  for the scalar variable $\zeta$ used in our toy model, the analogue of the strain measured at a detector located at $\bfx$ is obtained from the field operator $\hat\zeta(\bfx)$, whose matrix element in the field basis is diagonal~\cite{Colas:2023wxa},
\be
\langle \zeta^-_f|\hat\zeta(t_f, \bfx)|\zeta^+_f\rangle=\zeta^+_f(\bfx)\,\delta[\zeta_f^+-\zeta^-_f]\,.
\ee
Accordingly, the GW strain and the GW 2-point correlation function are defined 
\begin{align}
   \langle \hat \zeta(\bfx, t_f) \rangle_{\rm av} &\equiv {\rm Tr} \llp \hat \rho_{\rm av} (t_f) \, \hat \zeta (\bfx, t_f) \rrp = \int \dd \zeta^+_f \rho_{\rm av} (\zeta_f^+,\zeta_f^+;t_f) \zeta^+_f (\bfx) \label{eq:AverageZeta}\,, \\
    \langle \hat \zeta(\bfx, t_f) \hat \zeta(\bfy, t_f)  \rangle_{\rm av} &\equiv  {\rm Tr} \llp \hat \rho_{\rm av} \, \hat \zeta (\bfx, t_f) \hat \zeta (\bfy, t_f) \rrp =  \int \dd \zeta^+_f \rho_{\rm av} (\zeta_f^+,\zeta_f^+;t_f) \zeta^+_f (\bfx)  \zeta^+_f (\bfy) \label{eq:AverageZetaZeta}\,.
\end{align}
Working in Fourier basis, one can relate these quantities to the {\it tensor power spectrum}
\be 
{\rm Tr} \llp \rho_{\rm av}(t)\,\hat \zeta_{\bfk}\hat \zeta_{\bfk'} \rrp = \langle \hat \zeta_{\bfk}(t)\hat \zeta_{\bfk'}(t)\rangle_{\rm av} \equiv (2\pi)^3 \delta^{(3)}(\bfk+\bfk')\, P_h(k,t)\,,
\ee 
assuming statistical homogeneity and isotropy. 
This shows the explicit connection between the formalism that we will develop in this work and observables probed by interferometers.  For a stochastic GW background, these averaged quantities are directly related to the strain correlators measured by interferometers, while for resolved sources the same statistical information would have to be inferred from suitable two-point correlations constructed from lensing-sensitive observables across many events.
Using Eq.~\eqref{densitymatrixevolution}, we see indeed that the presence of matter structures, and thus gravitational potential wells, will affect the time evolved density matrix, thus generating an observable effect. 

\subsection{Disorder statistics}\label{sec:LensStat}

The final ingredient needed to characterize the system is the probability distribution functional of the gravitational potential. As a first step, we assume that gravitational potential is a Gaussian random field with vanishing mean. This is a well-motivated approximation on sufficiently large scales, where the statistics of cosmological perturbations are close to Gaussian consistently with the standard $\Lambda$CDM picture, while departures from Gaussianity become more relevant on small, non-linear scales \cite{Bartolo:2004if,Bertschinger:2001is,Bartolo:2005xa,Bernardeau:2001qr}. In the present work, our goal is to develop the formalism in the simplest setting, and we therefore restrict our attention to the Gaussian case.  

Accordingly, we take the probability distribution functional of the stochastic gravitational potential to be
\be
\label{Eq:Potential_distribution}
\mathcal{P}[\Phi] = {\cal N}_C\, {\rm exp}\llp -\frac12 \int \dd^3 \bfx \, \dd^3 \bfy \, \Phi(\bfx)\, C^{-1}(\bfx-\bfy)\,\Phi(\bfy) \rrp \,,
\ee
where ${\cal N}_C$ is a normalization constant fixed by the condition $\int \mathcal{D}\Phi \, \mathcal{P}[\Phi]=1$.
If one discretizes the spatial integrals into $N$ cells, the functional distribution reduces to an ordinary $N$-dimensional Gaussian, for which
\be
{\cal N}_C=\frac{1}{\sqrt{(2\pi)^N \det C}}\,.
\ee
In Eq.~\eqref{Eq:Potential_distribution}, $C^{-1}$ denotes the inverse two-point correlation function of the gravitational potential, defined by
\be
C(\bfx-\bfy) \equiv \int \mathcal{D}\Phi\,\mathcal{P}[\Phi]\, \Phi(\bfx)\Phi(\bfy)\,.
\ee

Under the assumption of statistical homogeneity, $C(\bfx-\bfy)$ depends only on the separation between the two points, while statistically isotropy would imply that $C(\bfx)$ depends exclusively on the modulus of the position $|\bfx|$. 
The Fourier transform $\tilde C$ of the two-point correlation function defines the {\it power spectrum} of the gravitational potential 
\be\label{eq:FTConventions}
C(\bfx) = \int \frac{\dd^3 \bfk}{(2\pi)^3} e^{i\bfk\cdot\bfx} \tilde C(\bfk)\,.
\ee
In cosmological applications, this quantity is related to the statistics of the gravitational potential generated by matter inhomogeneities. On large scales, where linear theory is valid, these are ultimately set by the primordial fluctuations, whose spectrum is known from CMB and large-scale-structure observations to be close to scale invariant. In the present analysis, $\tilde C(\bfk)$ will be kept general, so that different physical regimes can be incorporated through a suitable choice of the correlator. This allows the formalism to describe not only the statistics of standard cosmological perturbations, but also more general scenarios, including primordial black hole populations~\cite{Carr:2020gox} and dark-matter models with distinctive small-scale structure~\cite{Hu:2000ke,Hui:2016ltb}. 

The probability density functional can be written in Fourier space as well as 
\be \label{eq:PDFFT}
\P[\tilde\Phi]=  \tilde {\cal N}_C \exp\left\{-\frac12 \int \frac{\dd^3 \bfk}{(2\pi)^3} \: \tilde\Phi(\bfk) \: \tilde C^{-1}(\bfk) \: \tilde\Phi(-\bfk)\right\}\,,
\ee 
where $\tilde\Phi(\bfk)$ is the Fourier transformed gravitational potential, defined with the same conventions as in Eq.~\eqref{eq:FTConventions}.

\section{The time evolution operator}\label{sec:The time evolution operator}

From this section onward, we set $t_i=0$ and $t_f=T$, so that $T=t_f-t_i$. This choice is completely general in the present setup, since the gravitational potential is assumed to be time independent.

Our goal is to compute Eq.~\eqref{densitymatrixevolution}, for which we need the time evolution operator in Eq.~\eqref{Upath}. We will do so perturbatively, expanding in powers of the gravitational potential, however there are several initial steps that can be performed in full generality.

\subsection{Non-perturbative steps: background fields and loop corrections}

We employ the {\it background field method}, a standard technique in quantum field theory used to organize the path integral around a classical configuration and to compute fluctuation corrections systematically~\cite{Feynman:1963fq,Zinn-Justin:1989rgp,Schwartz:2014sze, Bessa:2007vq}. In the present case, this method can be applied particularly straightforwardly because the action is quadratic in the field.

We start by decomposing the GW field into a classical, {\it i.e.} on-shell, contribution and a fluctuation,
\be\label{eq:path separation}
\zeta(\bfx,t)= \zeta_{\rm cl}(\bfx,t)+\frac{1}{\sqrt{\Omega}}\,\eta(\bfx,t)\,,
\ee
where $\zeta_{\rm cl}(\bfx,t)$ satisfies the equation of motion~\eqref{eq:KG}. Since the path integral in Eq.~\eqref{Upath} is defined with fixed boundary conditions, Eq.~\eqref{eq:BC}, we assign them entirely to the classical part of the field
\be\label{eq:BCZetabar}
    \zeta_{\rm cl}(\bfx,t=0)= \zeta_i (\bfx)\,, \qquad \zeta_{\rm cl}(\bfx,t=T)= \zeta_f (\bfx)\,.
\ee
Accordingly, the fluctuation field vanishes at the initial and final times,
\be\label{Eq:Dirichelet_BC}
\eta(\bfx,t=0)=\eta(\bfx,t=T)=0\,,
\ee
so that it obeys Dirichlet boundary conditions.

The field $\eta$ in Eq.~\eqref{eq:path separation} describes fluctuations around the classical configuration and may also be viewed as the analog of a {\it Jacobi field} \cite{Schulman:1981vu}. In general, such fluctuations are treated perturbatively. Here, however, because the action is purely quadratic, the semiclassical expansion is exact and no higher-order terms are neglected.

Substituting the decomposition~\eqref{eq:path separation} into the action \eqref{eq:action in spacetime} yields
\begin{equation}\label{Eq:perturbed_action}
S[\zeta_{\rm cl}+\Omega^{-1/2}\eta,\Phi]
= S_{\rm cl}[\zeta_i,\zeta_f,\Phi] +\frac{1}{2\Omega}\int_0^T \dd^4x\,\dd^4y \; \eta(x)\,K_\Phi(x,y)\,\eta(y)\,,
\end{equation}
where the linear term vanishes because $\zeta_{\rm cl}$ is chosen to satisfy the equation of motion. In Eq.~\eqref{Eq:perturbed_action} we have introduced the following notation. First
\begin{equation}
S_{\rm cl}[\zeta_i,\zeta_f,\Phi] \equiv S[\zeta_{\rm cl},\Phi]
\end{equation}
denotes the action evaluated on the classical configuration. Note that this doesn't vanish for general boundary conditions, especially for finite propagation time. Once the equation of motion has been solved, $\zeta_{\rm cl}$ is completely determined by the boundary data and by the realization of the gravitational potential, which motivates the explicit notation $S_{\rm cl}[\zeta_i,\zeta_f,\Phi]$.
Then, the quadratic kernel is defined as 
\be 
\label{eq:Kphi_kernel}
K_\Phi(x,y) \equiv \llp -\partial_t^2+\partial_{\bfx}^2+4\alpha \Phi(\bfx)\partial_t^2 \rrp \delta^{(4)}(x-y)\,.
\ee
This is the inverse propagator of the fluctuation field. Because the original action is quadratic, the operator $K_\Phi$ depends on the external potential $\Phi$ but not on the background solution $\zeta_{\rm cl}$, and therefore not on the boundary data $\zeta_i$ and $\zeta_f$. This fact will be important later.

Performing the Gaussian integral, the time evolution operator then takes the form
\begin{align}\label{eq:semiclassical time evolution}
    \U_{\Phi} [\zeta_i, \zeta_f; 0, T]
    ={\cal N} \, \frac{e^{i\Omega S_{\rm cl}[\zeta_i,\zeta_f,\Phi] }}{\sqrt{\det{}^{\prime}(K_0^{-1}K_\Phi)}}\,,
\end{align}
where $\mathcal{N}$ is an overall normalization, and where we have introduced the free inverse propagator
\be
K_0(x,y)\equiv K_\Phi(x,y)\big|_{\Phi=0}\,.
\ee
Note that the ratio of determinants in Eq.~\eqref{eq:semiclassical time evolution}, representing the fluctuations around the classical trajectory, arises from the normalization of the Gaussian path integral and removes the corresponding free contribution. 
Dividing by $K_0^{-1}$ removes the background-independent vacuum normalization and isolates the effect of the external potential. Additionally, in Eq.~\eqref{eq:semiclassical time evolution} we have written $\det^{\prime}$ rather than $\det$. The prime indicates that the determinant is evaluated on the restricted space of fluctuation fields satisfying the Dirichlet boundary conditions~\eqref{Eq:Dirichelet_BC} (see~\cite{Zinn-Justin:1989rgp, Itzykson:1980rh}). In other words, the functional determinant is not taken over all field configurations, but only over those compatible with the boundary conditions of the path integral.

\bigskip
Eq.~\eqref{eq:semiclassical time evolution} is the main result of this part, and it is made of two contributions: $e^{i \Omega S_{\rm cl}}$ at the numerator, coming from the classical path contribution, and the determinants at the denominator accounting for the fluctuations. 
In the next sections we will work these out in perturbation theory at lowest orders.

\subsection{Evaluating the classical action}
\label{subsec:ClassicalAction}

In this subsection we compute the classical action evaluated on the saddle-point configuration, $S_{\rm cl}[\zeta_i,\zeta_f,\Phi]$. To do so, we first determine the classical field $\zeta_{\rm cl}(t,\bfx)$, which satisfies the equation of motion~\eqref{eq:KG} together with the boundary conditions~\eqref{eq:BCZetabar}. We solve this boundary-value problem perturbatively, expanding the classical solution in powers of the bookkeeping parameter $\alpha$ up to second order
\be
\label{Eq:perturbativeexpansionzeta}
\zeta_{\rm cl}=\zeta_0+\alpha\,\zeta_1+\alpha^2\,\zeta_2\,.
\ee
We keep terms up to $\mathcal O(\alpha^2)$ in order to match the lowest nontrivial order of the fluctuation determinant (see next Section), which also starts at second order in $\alpha$. \footnote{There are additional contributions at order $\alpha^2$ that would arise by expanding Eq.~\eqref{eq:KGgeneral} itself to second order in the gravitational potential. In the present work we neglect these terms for simplicity. They would schematically generate corrections of the form $\alpha^2\Phi^2\partial_t^2$ and $\alpha^2\partial^i\Phi\,\partial_i\Phi$. The second type is structurally different and would have to be neglected on the basis of an order-of-magnitude argument, for instance by assuming that spatial gradients of the potential are small compared to the frequency of the wave. The first type could instead be absorbed into a redefinition $\Phi\to \Phi+\alpha\Phi^2$, but such a redefinition would modify the statistical properties of the random field and in particular spoil the Gaussian assumption. We therefore leave these effects for future work.}

Substituting the perturbative ansatz into Eq.~\eqref{eq:KG} and matching order by order in \(\alpha\), we obtain
\be
\Box\,\zeta_0 = 0 \,, \qquad
\Box\,\zeta_1 = -4\,\Phi\,\partial_t^2 \zeta_0 \,, \qquad
\Box\,\zeta_2 = -4\,\Phi\,\partial_t^2 \zeta_1 \,,
\ee
where $\Box =  -\partial_t^2+\nabla^2$ is the free wave operator on Minkowski spacetime. We assign the prescribed boundary data to the zeroth-order solution,
\be\label{eq:BCzeta0}
\zeta_0(\bfx,t= 0)=\zeta_i(\bfx)\,, \qquad
\zeta_0(\bfx,t= T)=\zeta_f(\bfx)\,,
\ee
while the higher-order corrections vanish at the endpoints,
\be\label{eq:BCzeta12}
\zeta_1(\bfx,t= 0)=\zeta_1(\bfx,t= T)=0\,, \qquad
\zeta_2(\bfx,t= 0)=\zeta_2(\bfx,t= T)=0\,.
\ee
This choice ensures that the full classical solution $\zeta_{\rm cl}$ satisfies the required boundary conditions order by order in $\alpha$.
The zeroth-order field is therefore the free solution interpolating between the two boundary configurations, and can be easily found. The higher-order corrections can be obtained by means of the free Green function  as 
\begin{align}
\zeta_1(x) &= -4\int_0^T \dd^4 x' \:
\Delta_0(x,x') \: \Phi(\bfx') \:  \partial_{t'}^2\zeta_0(x')\,, \label{eq:zeta1FromGreen} \\
\zeta_2(x) &= 16\int_0^T \dd^4 x'\,\dd^4 x'' \: \Delta_0(x,x') \partial_{t'}^2 \Delta_0(x',x'') \: \Phi(\bfx')  \Phi(\bfx'') \: \partial_{t''}^2 \zeta_0(x'') \,, \label{eq:zeta2FromGreen}
\end{align}
where we have used the notation $x' = (t', \bfx')$ and $x'' = (t'', \bfx'')$.

In these expressions, $\Delta_0$ is the free propagator, defined by
\be\label{eq:FreePropDef}
\int d^4x'\,K_0(x,x')\,\Delta_0(x',y)= \delta^{(4)}(x-y)\,,
\ee
with Dirichlet boundary conditions at $t=0$ and $t=T$, so that the boundary conditions of $\zeta_1$ and $\zeta_2$ are automatically implemented. Since the free problem is translationally invariant in space, the Green function depends only on the spatial separation,
\be
\Delta_0(x,y)=\Delta_0(\bfx-\bfy,t_x,t_y)\,,
\ee
and satisfies
\be
\Delta_0(\bfx-\bfy,0,t_y)=\Delta_0(\bfx-\bfy,T,t_y)=0\,,
\ee
with analogous conditions when $t_y=0$ or $t_y=T$.
Because the propagation takes place over the finite time interval $[0,T]$, the temporal dependence is naturally expanded in the discrete set of normal modes compatible with these boundary conditions. It is therefore convenient to solve for the propagator in Fourier space, where one obtains~\cite{Bellac:2011kqa}
\be \label{eq:FreeGFkt} 
\tilde \Delta_0 (\bfk,t_x,t_y) = \sum_{n=1}^{\infty}\frac{2}{T} \frac{\sin(\omega_n t_x)\sin(\omega_n t_y)}{\omega_n^2-|\bfk|^2}\,, \qquad \omega_n=\frac{\pi n}{T}\,.
\ee
Resumming the series then gives the equivalent closed expression~\cite{Bessa:2007vq}
\be \label{Eq:free_prop_resummed} 
\tilde \Delta_0 (\bfk,t_x,t_y) = \theta(t_x-t_y)\, \frac{\sin(|\bfk| t_y)\,\sin[|\bfk|(T-t_x)]}{|\bfk|\sin(|\bfk| T)} + \theta(t_y-t_x)\,\frac{\sin(|\bfk| t_x)\,\sin[|\bfk|(T-t_y)]}{|\bfk|\sin(|\bfk| T)}\,.
\ee

Note that $\{  \zeta_1,  \zeta_2 \}$ still depend on the initial and final field configurations $\{ \zeta_i, \zeta_f \}$, as their source is built out of the free solution, which is completely determined by those boundary values. Substituting the perturbative expansion \eqref{Eq:perturbativeexpansionzeta} into the action yields
\be \label{actionclassical}
S_{\rm cl} [\zeta_i, \zeta_f, \Phi] = S_0[ \zeta_i,\zeta_f ] +\alpha S_1[ \zeta_i,\zeta_f,\Phi ]  +\alpha^2 S_2[ \zeta_i,\zeta_f,\Phi ] \,,
\ee 
where the explicit expressions for $S_0$, $S_1$, and $S_2$ are computed inserting the explicit forms of $\zeta_0$, $\zeta_1$ and $\zeta_2$ into the action. Their explicit forms are computed in the Appendix~\ref{App:A}, and are given in Eqs.~\eqref{eq:S0classical},~\eqref{eq:S1classicalFourier} and~\eqref{eq:S2classicalFourier}.

\subsection{Loop Contributions}
\label{subsec:Loops}
\begin{figure}
    \centering
    \includegraphics[width=0.8\linewidth]{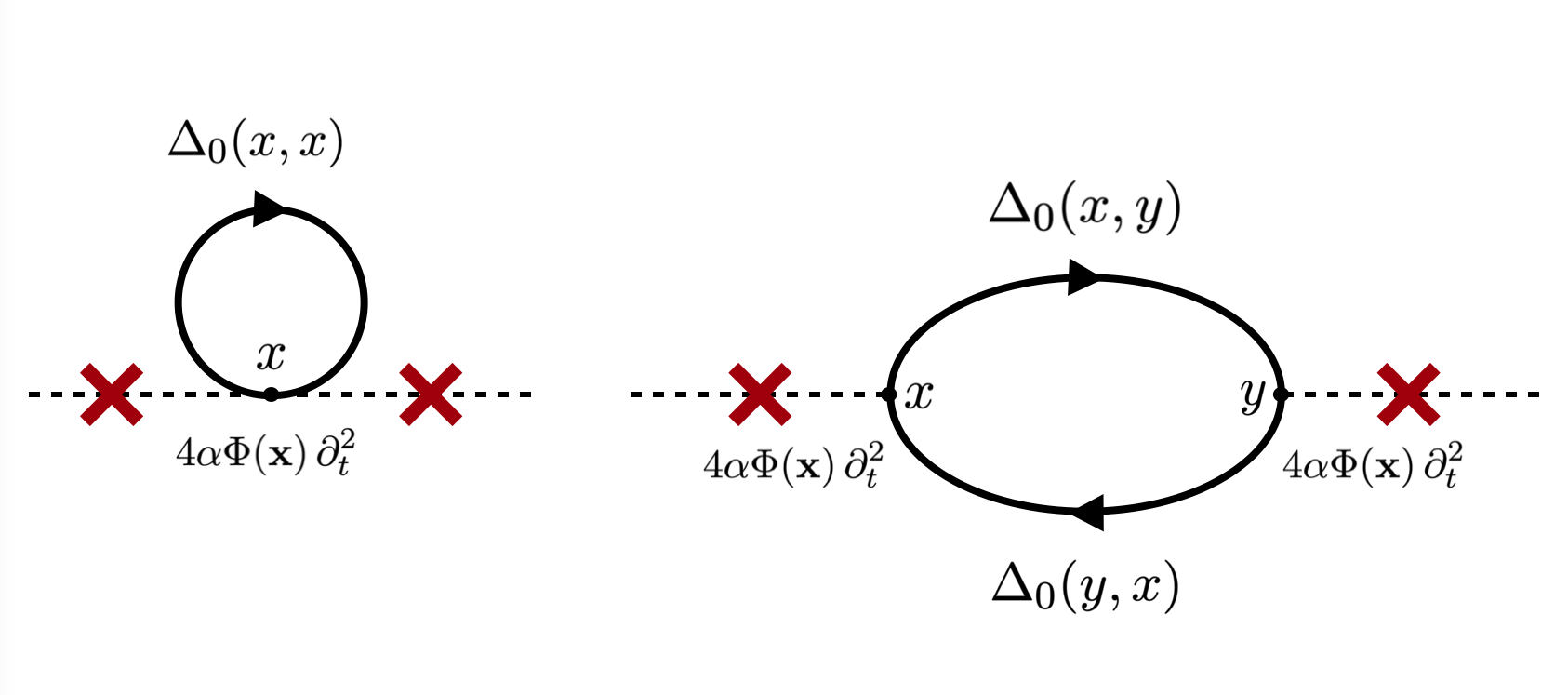}
    \caption{Loop expansion of the time evolution operator.}
    \label{fig:LoopsContributions}
\end{figure}

We now turn to the denominator in Eq.~\eqref{eq:semiclassical time evolution}, which encodes the contribution of fluctuations around the saddle point. Since the action is quadratic in the field $\zeta$, the path integral over fluctuations is Gaussian and gives rise to a functional determinant. This contribution can be written as
\begin{align}
    \label{eq:semiclassical_determinat}
    \frac{1}{\sqrt{(\det^{\prime}K_0^{-1}K)}} 
    &= \exp\left[-\frac{1}{2} {\rm Tr} \ln \left( 1 + K_0^{-1} \circ \delta K \right)\right]\,,
\end{align}
where $\delta K = K_\Phi - K_0$ is perturbed part of the propagator. The trace of an operator $\hat A$ is defined as
\be \label{eq:Trace}
{\rm Tr}\,\hat A \equiv \int_0^T \dd^4x\, A(x,x)\,,
\qquad
A(x,y)=\langle x|\hat A|y\rangle \,.
\ee
Working in the weak-potential regime, we expand the logarithm in Eq.~\eqref{eq:semiclassical_determinat} perturbatively and truncate at second order. In this way the fluctuation contribution becomes
\begin{align}\label{Eq:Loop corrections}
    &\frac{1}{\sqrt{(\det^{\prime}K_0^{-1}K)}} = \nonumber \\
    &~~ =\exp\left\{-\frac{1}{2} {\rm Tr}\llp 4\alpha \Phi({\bfx})\, \partial_{t_x}^2 \Delta_0(x, y)-8\alpha^2 \int_{0}^{T} \dd^4 x'\, \Phi({\bfx})\, \partial_{t_x}^2 \Delta_0(x, x')\, \Phi(\bfx')\, \partial_{t_{x'}}^2 \Delta_0(x', y)  \rrp\right\}\,,
\end{align}
where $x=(t_x,\bfx)$, $x'=(t_{x'},\bfx')$.
These two contributions are represented diagrammatically in Fig.~\ref{fig:LoopsContributions}. 
Using the explicit expression for the propagator in Eq.~\eqref{eq:FreeGFkt}, the fluctuation determinant can be written as
\begin{align}
\label{eq:dressed_TEO_2order}
   \frac{1}{\sqrt{(\det^{\prime}K_0^{-1}K)}} &= \exp\bigg\{2\alpha \Tilde{\Phi}(0)\int \frac{\dd^3 \bfk}{(2\pi)^3}\sum_{n=1}^{\infty}\frac{\omega_n^2}{\omega_n^2-|\bfk|^2}+\nonumber\\
   &\qquad +4\alpha^2\int \frac{\dd^3 \bfq\, \dd^3 \bfk}{(2\pi)^6}\Tilde{\Phi}( \bfq) \Tilde{\Phi}(- \bfq )\sum_{n=1}^{+\infty}\frac{\omega_n^4 }{[\omega_n^2-|\bfk|^2][\omega_n^2-|({\bfk}+\boldsymbol{q})|^2]}\bigg\}\,,
\end{align}
with  $\tilde\Phi$ the Fourier transform of the gravitational potential.
The details of this computation are reported in Appendix~\ref{App:B}.  
The term linear in $\alpha$  in Eq.~\eqref{eq:dressed_TEO_2order} corresponds to a {\it tadpole} contribution~\cite{Schulman:1981vu,Schwartz:2014sze,Peskin:1995ev}. The appearance of the Fourier transform evaluated in $0$ shows explicitly that this term probes only the constant part of the background, namely the spatial average of the disorder, indeed
\be 
\Tilde{\Phi}(0) \sim \int d^3\mathbf x\,\Phi(\mathbf x)\,.
\ee 
Assuming ergodicity, one could relate this spatial average with the ensemble average of the gravitational potential
\be 
\int d^3\mathbf x\,\Phi(\mathbf x) \sim \int {\cal D} \Phi \, \P[\Phi] \Phi\,,
\ee 
for large enough volumes.  Since in our stochastic setup the disorder is assumed to have vanishing mean (see Sec.~\ref{sec:LensStat}), the linear correction vanishes identically after averaging. For this reason, in the presence of zero-mean disorder the leading non-vanishing contribution arises at order $\alpha^2$. 

The second term in Eq.~\eqref{eq:dressed_TEO_2order}, gives the leading non-trivial correction to the time evolution operator. Since the action in Eq.~\eqref{eq:action in spacetime} is purely quadratic in the fluctuations, the path integral over $\eta$ is Gaussian and no interaction vertices involving the dynamical scalar field are present. Consequently, for a fixed realization of the disorder, there are no radiative self-interactions of $\zeta$ beyond the exact Gaussian propagator determined by the quadratic operator. Because this operator depends on the external random field $\Phi(\bfx)$, integrating out the dynamical field generates a non-trivial one-loop effective action for the disorder. In this sense, the quantum fluctuations of $\zeta$ induce an infinite series of generally nonlocal terms built out of $\Phi(\bfx)$~\cite{Schwartz:2014sze,Zinn-Justin:1989rgp,Feynman:1963fq,Schulman:1981vu}, of the type represented in Fig.~\ref{fig:LoopsContributions}, where each insertion can be interpreted as an interaction with the gravitational potential. Thus, although $\zeta$ remains a free field in a fixed background, its fluctuations generate non-trivial contributions to the effective statistical weight of the random disorder (see Fig.~\ref{fig:quantum_corrected_PHI}).

We note that, since the propagation takes place over a finite time interval, the temporal spectrum is discrete and the loop contribution is naturally written as a sum over modes. In this setting, the expression in Eq.~\eqref{eq:dressed_TEO_2order} is well-defined without specifying any pole prescription. A prescription becomes necessary only in the limit $T\to +\infty$, where the discrete frequencies become dense, the mode sum reorganizes into the usual continuum representation, and one recovers the standard singular structure of the infinite-time Green function. As we will show below, however, only the real part of Eq.~\eqref{eq:dressed_TEO_2order} is relevant for our purposes, since our final results will depend only on the square modulus of this contribution.

The two sums in Eq.~\eqref{eq:dressed_TEO_2order} can be resummed
\begin{align}
    \sum_{n=1}^{+\infty}  \frac{n^2}{(n^2 - a^2)} &= \frac12 - \frac{a \pi}{2} \cot(\pi a) + \sum_{n=1}^{+\infty}1 \,, \\
    \sum_{n=1}^{+\infty}  \frac{n^4}{(n^2 - a^2)(n^2 - b^2)} &= \frac12 + \frac{b^3\pi\cot(\pi b)-a^3\pi\cot(\pi a)}{2(a^2-b^2)} + \sum_{n=1}^{+\infty}1 \,,
\end{align}
where we have used the Mittag-Leffler~\cite{10.5555/1098650} expansion for cotangent function. This resummation makes explicit the diverging factor (last contribution), which however can be reabsorbed in the normalization of the time evolution operator. 
The resummed version of Eq.~\eqref{eq:dressed_TEO_2order},  obtained using these identities, coincides with the expression of the loop correction computed with the resummed propagator of Eq.~\eqref{Eq:free_prop_resummed}.

Finally, we stress that neither of these two terms in Eq.~\eqref{eq:dressed_TEO_2order} depends on the boundary field configurations ($\zeta_i, \zeta_f$), since they arise entirely from the Gaussian integral over the Jacobi field, which obey Dirichlet boundary conditions. Another consequence of this fact is that this loop contribution is independent of semiclassical parameter $\Omega$ as well, so that this term survives in any regime of $\Omega$.

\section{Evolution of the averaged density matrix}\label{sec:Density matrix results}

Using the semiclassical form of the time-evolution operator derived in Eq.~\eqref{eq:semiclassical time evolution}, the averaged density matrix at time $t$ can be written as
\begin{align}\label{eq:RhoTimeEvolved}
    \rho_{\rm av}(\zeta_f^+,\zeta_f^-;t) &= {\cal N} \, \int \mathcal D\Phi\, \frac{\P[\Phi]}{\Big|\sqrt{\det{}^{\prime}(K_0^{-1}K_\Phi)} \Big|^2} \, \times \nonumber \\
    &~~\times \int \dd \zeta_i^+ \dd \zeta_i^-  \: \rho(\zeta_i^+,\zeta_i^-;t_i) \: \exp \lp i\Omega  \lp  S_{\rm cl}[\zeta^+_i,\zeta^+_f,\Phi] -  S_{\rm cl}[\zeta^-_i,\zeta^-_f,\Phi] \rp \rp  \,.
\end{align}
To arrive to this form, we have used that the loop correction determinant factor doesn't depend on the initial field configurations on the two branches, thus it can be pulled out of the corresponding integral. We have also introduced the compact notation
\begin{align}\label{eq:ActionOnShellDifference}
 S_{\rm cl} [\zeta^+_i,\zeta^+_f,\Phi] -  S_{\rm cl}[\zeta^-_i,\zeta^-_f,\Phi] &\equiv \delta  S_{\rm cl}[\zeta^\pm_i,\zeta^\pm_f,\Phi] \nonumber\\
& \equiv \delta S_0[\zeta^\pm_i,\zeta^\pm_f] + \alpha \,\delta S_1[\zeta^\pm_i,\zeta^\pm_f,\Phi] + \alpha^2 \,\delta S_2[\zeta^\pm_i,\zeta^\pm_f,\Phi]\,,
\end{align}
where the difference is always understood as the $(+)$ branch minus the $(-)$ branch, and $\delta S_i$ are defined as the components of $\delta  S_{\rm cl}$ in a perturbative expansion in powers of the gravitational potential at order $\alpha^i$.
Finally, the normalization constant ${\cal N}$ is fixed by imposing trace preservation,
\be
{\rm Tr}\,\llp \rho_{\rm av}(t) \rrp =\int \dd \zeta_f \, \rho_{\rm av}(\zeta_f,\zeta_f;t)=1\,.
\ee
All the ingredients entering Eq.~\eqref{eq:RhoTimeEvolved} were computed in Secs.~\ref{subsec:ClassicalAction} and~\ref{subsec:Loops}. In this section we combine them in order to determine the structure of the time-evolved averaged density matrix.

\subsection{Effective gravitational potential power spectrum}
Eq.~\eqref{eq:RhoTimeEvolved} makes transparent that the role of loop determinant  is to modify the statistical weight with which each realization of the disorder is weighted, by reshaping the probability density functional of the disorder into the effective one
\be \label{eq:Peff}
{\cal P}^{\rm eff}\llp \Phi \rrp \equiv \frac{{\cal P}\llp \Phi \rrp}{\Big|\sqrt{\det{}^{\prime}(K_0^{-1}K_\Phi)} \Big|^2}\,.
\ee
Since this contribution is branch-independent, it cannot generate loss of phase coherence, as the underlying mechanism of these processes is controlled by the phase mismatch between the two Schwinger--Keldysh branches.  Thus, they affect the non-unitary dynamics only indirectly, for instance by changing the quantitative efficiency of such processes.
Using Eqs.~\eqref{eq:dressed_TEO_2order}\footnote{The linear term in Eq.~\eqref{eq:dressed_TEO_2order} can be neglected under the assumptions discussed in Sec.~\ref{subsec:Loops}; if retained, it would correspond to a shift of the mean of the Gaussian measure rather than to a modification of its covariance.}, one can then write
\be \label{eq:EffectivePDF}
\P^{\rm eff} [\Phi] = \tilde{\cal N}_{\rm eff} \exp \left\{-\frac12 \int \frac{\dd^3 \bfk}{(2 \pi)^3} \: \tilde\Phi(\bfk) \: \tilde C^{-1}_{\rm eff}(\bfk) \: \tilde\Phi(-\bfk)\right\}\,,
\ee
where the inverse of the effective power spectrum is
\be \label{eq:EffectivePS}
\tilde C^{-1}_{\rm eff}(\bfk)  = \tilde C^{-1}(\bfk) - 16 \alpha^2 {\rm Re} \llp  \int \frac{\dd^3 \bfq}{(2\pi)^3} \sum_{n=1}^{+\infty}\frac{\omega_n^4 }{[\omega_n^2-|\bfq|^2][\omega_n^2-( \bfk+ \bfq)^2]} \rrp \,.
\ee
Here $\tilde{\cal N}_{\rm eff}$ denotes the corresponding normalization constant, which differs from the original one once the loop correction is included. Eq.~\eqref{eq:EffectivePS} exhibits an ultraviolet divergence characteristic of loop contributions, which has to be renormalized. We leave the details of this computation to future work, here we simply regularize with a cutoff the divergent integrals.

\begin{figure}[H]
    \centering
    \includegraphics[width=0.8\linewidth]{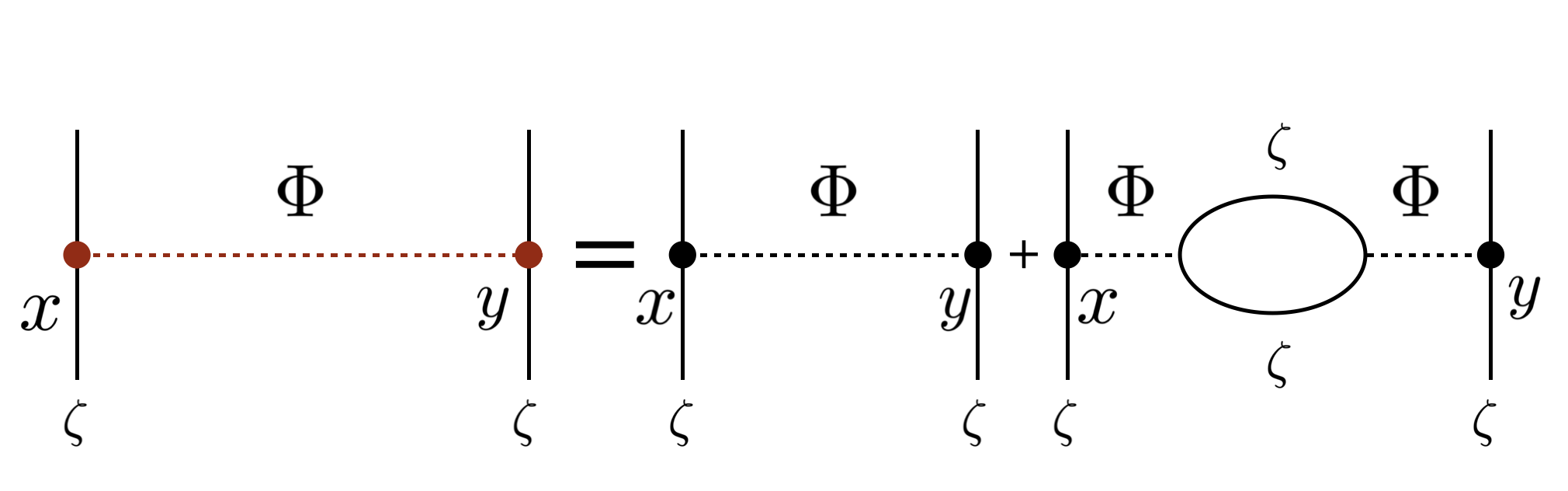}
    \caption{Diagrammatic description of the dressed disorder correlator}
    \label{fig:quantum_corrected_PHI}
\end{figure}

Equations~\eqref{eq:EffectivePDF} and~\eqref{eq:EffectivePS} make explicit the interpretation anticipated in Sec.~\ref{subsec:Loops}. Integrating out the fluctuations of $\zeta$ produces a loop correction to the statistical weight of the gravitational potential. The disorder ensemble is renormalized, and the two-point function of the background field is modified according to the effective kernel $\tilde C_{\rm eff}^{-1}$.
Thus, the fluctuations of the scalar field induce a non-trivial correction to the effective distribution of lens configurations entering the averaged density matrix (see Fig.~\ref{fig:quantum_corrected_PHI}).

\subsection{Diagonal elements}\label{sec:DiagonalElements}

We begin from Eq.~\eqref{eq:RhoTimeEvolved} and restrict it to the diagonal sector, namely $\zeta_f^+=\zeta_f^- \equiv \zeta_f$. As previously described, these components are important for the averages of operators which are diagonal in the $\zeta$-basis (see Eqs.~\eqref{eq:AverageZeta} and~\eqref{eq:AverageZetaZeta}). 
It is instructive to consider first the special case in which the initial density matrix is itself diagonal in the $\zeta$-basis,
\be
\rho(\zeta_i^+,\zeta_i^-;t_i)=\rho(\zeta_i^+,\zeta_i^+;t_i)\,\delta(\zeta_i^+-\zeta_i^-)\,.
\ee
In this case the delta function forces the two Schwinger--Keldysh branches to coincide already at the initial time, so that the classical phase difference vanishes identically once the final diagonal condition $\zeta_f^+=\zeta_f^-$ is imposed. The averaged diagonal element then reduces to
\be
\rho_{\rm av}(\zeta_f,\zeta_f;t_f) = {\cal N}\int \mathcal D\Phi\,\P^{\rm eff}[\Phi]\int d\zeta_i\,\rho(\zeta_i,\zeta_i;t_i) = {\cal N}\,,
\ee
where in the last step we used the normalization of both the effective disorder distribution and the initial density matrix. This shows that, if the initial states are uncorrelated, the final diagonal elements become insensitive to the disorder, precisely because the branch coherence has already been removed at the initial time. 

From a quantum field theory perspective, this result comes from considering a purely quadratic generating functional, with no sources. It is well known~\cite{Zinn-Justin:1989rgp,Schwartz:2014sze} that the sourceless ($J=0$) generating functional of the connected diagrams only generates vacuum amplitude contributions, which can always be reabsorbed into the normalization. In diagrammatic terms, Wick’s theorem~\cite{PhysRev.80.268} implies that all contractions of fields in a Gaussian path integral without sources produce closed loops with no external legs, corresponding to vacuum bubbles. These contributions are independent of the external configurations and can therefore be absorbed into the overall normalization.
On the other hand, if the initial density matrix is not diagonal, the dependence on the initial field configurations (different on the two in-in branches) effectively acts as sources, keeping the path integral sensitive to nontrivial correlations. By restricting attention to the diagonal elements, we are effectively “turning off” these sources and focusing only on the vacuum amplitudes, which justifies interpreting the contribution as a normalization factor.

Keeping the initial density matrix completely general, then $\delta  S_{\rm cl} [\zeta^\pm_i, \zeta_f, \Phi]\neq 0$. This is already clear at zeroth order, where one finds
\begin{align}\label{eq:DeltaS0PM}
&\delta S_0[\zeta^\pm_i,\zeta_f] = S_0[\zeta^+_i,\zeta_f] - S_0[\zeta^-_i,\zeta_f] \nonumber\\
&~~= \int \frac{\dd^3 \bfk}{(2\pi)^3} \frac{|\bfk|^2 T}{\sin \lp |\bfk| T\rp}
\llp  |\zeta^{i+}_\bfk|^2 - |\zeta^{i-}_\bfk|^2  - \cos (|\bfk| T) \llp \zeta^f_{-\bfk} \lp \zeta^{i+}_\bfk  - \zeta^{i-}_\bfk  \rp + \zeta^f_{\bfk} \lp \zeta^{i+}_{-\bfk}  - \zeta^{i-}_{-\bfk}  \rp \rrp \rrp\,,
\end{align}
where we used the explicit expression of $S_0[\zeta_i,\zeta_f]$ given in Eq.~\eqref{eq:S0classical} of Appendix~\ref{AppA:free}. This expression shows explicitly that the behavior of $\delta S_0[\zeta^\pm_i,\zeta_f]$ is controlled by the mismatch between the initial field configurations on the two Schwinger--Keldysh branches.
The same conclusion holds for $\delta S_1[\zeta^\pm_i,\zeta_f,\Phi]$ and $\delta S_2[\zeta^\pm_i,\zeta_f,\Phi]$, although their explicit expressions are more involved (see Appendix~\ref{AppA:first} and~\ref{AppA:second}).

At this stage one could continue the analysis by writing the diagonal sector explicitly, but this leads to lengthy expressions that are no simpler than those appearing in the general case $\zeta^+_f \neq \zeta^-_f$. For this reason, in the next subsection we proceed directly with the computation of the full density matrix, and recover the diagonal sector by imposing $\zeta_f^+=\zeta_f^-$ only at the end.

\subsection{Off-diagonal elements}

We now continue with the derivation for generic density-matrix elements. 

Using Eqs.~\eqref{eq:S0classical},~\eqref{eq:S1classicalFourier} and~\eqref{eq:S2classicalFourier}, we write the on-shell action as
\be
 S_{\rm cl} [\zeta_i, \zeta_f,\Phi] = S_0[\zeta_i, \zeta_f] + \int \frac{\dd^3 \bfk}{(2 \pi)^3} \: \tilde{\Phi}(\bfk) \: {\cal A}_1 \lp \bfk, \zeta_i, \zeta_f\rp + \int \frac{\dd^3 \bfk \dd^3 \bfq}{(2 \pi)^6} \: \tilde{\Phi}(\bfk) \tilde{\Phi}(\bfq ) \:  {\cal A}_2 \lp \bfk, \bfq, \zeta_i, \zeta_f\rp\,,
\ee
where ${\cal A}_1$ and ${\cal A}_2$ are given in Eqs.~\eqref{eq:A1Fourier} and~\eqref{eq:A2Fourier}, respectively. Inserting this into Eq.~\eqref{eq:ActionOnShellDifference}, we obtain
\begin{align}
\delta S_1[\zeta^\pm_i,\zeta^\pm_f,\Phi]
&= \int \frac{\dd^3 \bfk}{(2 \pi)^3} \: \tilde{\Phi}(\bfk) \: \times \: \delta {\cal A}_1 [\bfk, \zeta^\pm_i,\zeta^\pm_f]\,, \\
\delta S_2[\zeta^\pm_i,\zeta^\pm_f,\Phi]
&= \int  \frac{\dd^3 \bfk \dd^3 \bfq}{(2 \pi)^6} \: \tilde{\Phi}(\bfk) \tilde{\Phi}(\bfq ) \: \times \:  \delta {\cal A}_2 [\bfk, \bfq , \zeta^\pm_i,\zeta^\pm_f]\,,
\end{align}
with
\begin{align}
\delta {\cal A}_1 [\bfk, \zeta^\pm_i,\zeta^\pm_f]
&\equiv {\cal A}_1 \lp \bfk, \zeta^+_i, \zeta^+_f\rp - {\cal A}_1 \lp \bfk, \zeta^-_i, \zeta^-_f\rp\,, \\
\delta {\cal A}_2 [\bfk, \bfq , \zeta^\pm_i,\zeta^\pm_f]
&\equiv {\cal A}_2 \lp \bfk, \bfq, \zeta^+_i, \zeta^+_f\rp  -  {\cal A}_2 \lp \bfk, \bfq, \zeta^-_i, \zeta^-_f\rp\,.
\end{align}
The same convention is adopted throughout: the difference is always taken as the $(+)$ branch minus the $(-)$ branch.
Eq.~\eqref{eq:RhoTimeEvolved} can then be rewritten as
\begin{align}
&\rho_{\rm av}(\zeta_f^+,\zeta_f^-;t) = {\cal N} \, \int \dd \zeta_i^+ \dd \zeta_i^- \: \rho(\zeta_i^+,\zeta_i^-;t_i) \: e^{i\Omega \delta S_0[\zeta^\pm_i,\zeta^\pm_f] } \nonumber\\
&~~~~~ \times \int \mathcal D\Phi\, \exp \Bigg\{
i \alpha \Omega \int \frac{\dd^3 \bfk}{(2 \pi)^3} \: \tilde{\Phi}(\bfk) \, \delta {\cal A}_1 [\bfk, \zeta^\pm_i,\zeta^\pm_f] +
\nonumber\\
&~~~~~~~~ -\frac12 \int  \frac{\dd^3 \bfk \dd^3 \bfq}{(2 \pi)^6} \tilde{\Phi}(\bfk) \llp \tilde C^{-1}_{\rm eff}(\bfk) (2\pi)^3 \delta^{(3)} (\bfk + \bfq) - 2 i \alpha^2 \Omega \,\delta {\cal A}_2 [\bfk, \bfq , \zeta^\pm_i,\zeta^\pm_f]  \rrp \tilde{\Phi} (\bfq ) \Bigg\}\,,
\end{align}
where we have used Eq.~\eqref{eq:EffectivePDF} and assumed that the order of integration can be exchanged.
Performing the Gaussian path integral yields
\begin{align}
& \rho_{\rm av}(\zeta_f^+,\zeta_f^-;t) = {\cal N} \int \dd \zeta_i^+ \dd \zeta_i^- \: \rho(\zeta_i^+,\zeta_i^-;t_i)\: e^{i\Omega \delta S_0[\zeta^\pm_i,\zeta^\pm_f]}  \nonumber\\
&~~~~~ \times \frac{1}{\sqrt{\det {\cal M}}} \exp\Bigg\{ -\frac{\alpha^2 \Omega^2}{2} \int \frac{\dd^3 \bfk \dd^3 \bfq}{(2\pi)^6}\delta{\cal A}_1[\bfk,\zeta_i^\pm,\zeta_f^\pm]\,{\cal M}^{-1}(\bfk,\bfq;\zeta_i^\pm,\zeta_f^\pm)\,\delta {\cal A}_1[\bfq,\zeta_i^\pm,\zeta_f^\pm]\Bigg\}\,,
\label{eq:RhoAvAfterGaussian}
\end{align}
where we have defined the new kernel
\be
\label{eq:KernelM}
{\cal M}(\bfk,\bfq;\zeta_i^\pm,\zeta_f^\pm) \equiv \tilde C^{-1}_{\rm eff}(\bfk)\,(2\pi)^3 \delta^{(3)}(\bfk+\bfq) -2i \alpha^2 \Omega\,\delta{\cal A}_2[\bfk,\bfq,\zeta_i^\pm,\zeta_f^\pm]\,.
\ee
Eq.~\eqref{eq:RhoAvAfterGaussian} is the explicit result of the disorder average and provides a compact representation of how the stochastic lens distribution modifies the propagation of the wave. Its structure is naturally separated into three parts. The factor $\exp(i\Omega \delta S_0[\zeta_i^\pm,\zeta_f^\pm])$ is the free relative phase between the two Schwinger--Keldysh branches and is the remnant of the unitary evolution in the absence of disorder. 
The factor $(\det {\cal M})^{-1/2}$ arises from performing the Gaussian functional integral over the stochastic potential and modifies the statistical weight of different branch configurations. The exponential quadratic form built from $\delta{\cal A}_1$ and ${\cal M}^{-1}$ captures the dependence of the averaged evolution on the branch mismatch induced by the disorder. 
Note that the new kernel ${\cal M}^{-1}(\mathbf k,\mathbf q, \zeta^\pm_i, \zeta^\pm_f)$ is both non-local in momentum space and it depends on the boundary fields' configurations because of the contribution coming from the quadratic expansion of the action evaluated on the on-shell trajectory, i.e. $\delta {\cal A}_2$. This means that different Fourier modes are coupled due to disorder-averaging and, consequently, the suppression of a given density-matrix element is not controlled mode by mode independently, but by a correlated combination of momentum modes weighted by the effective kernel.

The dependence on the parameter $\Omega$ is also physically instructive. For small $\Omega$, both the oscillatory factor $\exp(i\Omega \delta S_0)$ and the exponential suppression term become weak. Moreover, ${\rm det}\,{\cal M}$ becomes branch independent and can therefore be absorbed into the overall normalization of the density matrix. In this limit, the coupling to the static gravitational potential becomes ineffective, and the propagation is insensitive to the disorder.

By contrast, in the $\Omega\to\infty$ limit the complex phases in Eq.~\eqref{eq:RhoAvAfterGaussian} become rapidly oscillatory, while the quadratic exponential becomes increasingly selective with respect to branch configurations for which $\delta{\cal A}_1\neq 0$. As a result, the integral is dominated by stationary-phase configurations, and branch mismatches that are not supported by the saddle structure are strongly suppressed.

These considerations give a clear physical interpretation of $\Omega$ as the frequency of the wave through the correspondence principle. Since the gravitational potential is static, Eq.~\eqref{eq:KG} is invariant under time translations, and its solutions can be decomposed into modes of the form $\zeta(t,\mathbf{x})=e^{i\omega t}\zeta_\omega(\mathbf{x})$, where $\omega$ is the frequency. In the semiclassical path-integral formulation, the free propagation is weighted by the phase $e^{i\Omega S_0}$, so that $\Omega$ plays precisely the role of the parameter controlling the oscillatory behavior of the mode. By the correspondence principle, one is thus led to identify $\Omega$ with $\omega$. This is consistent with the structure of Eq.~\eqref{eq:KG}, where the coupling to the stochastic gravitational potential is through time derivatives of the field: in the low-$\Omega$ regime the wave varies slowly in time and the interaction with the background is correspondingly weak, whereas for large $\Omega$ the dynamics approaches the short-wavelength semiclassical regime dominated by stationary phases~\cite{2010EJPh...31..171M,Schulman:1981vu,goldstein2002classical}.

\subsection{Decoherence}
In this section we take one step further from Eq.~\eqref{eq:RhoAvAfterGaussian}, in order to obtain an explicit expression in which oscillatory phase factors and real exponential suppression terms are clearly separated. Exploiting the smallness of $\alpha$, we expand the last determinant appearing in Eq.~\eqref{eq:RhoAvAfterGaussian}. 
To this end, we isolate from ${\cal M}$ the part that is independent of the boundary fields on the two branches. Defining
\be 
{\cal M}_0(\bfk, \bfq) = \tilde C^{-1}_{\rm eff}(\bfk)\,(2\pi)^3 \delta^{(3)}(\bfk+\bfq) \,,  \qquad \mbox{with} \qquad {\cal M}^{-1}_0(\bfk, \bfq) = \tilde C_{\rm eff}(\bfk)\,(2\pi)^3 \delta^{(3)}(\bfk+\bfq)\,,
\ee 
we expand the determinant as 
\begin{align} 
\frac{1}{\sqrt{\det {\cal M}}} &= \frac{1}{\sqrt{\det {\cal M}_0}} \exp \bigg\{ -\frac{1}{2} {\rm Tr} \ln  \lp 1 - 2 i \alpha^2 \Omega\, \llp {\cal M}_0^{-1} \circ \,\delta{\cal A}_2 [\zeta_i^\pm,\zeta_f^\pm] \rrp_{\bfk, \bfq} \rp   \bigg\} = \\
&\approx \frac{1}{\sqrt{\det {\cal M}_0}} \exp \bigg\{  i \alpha^2 \Omega\,  {\rm Tr} \llp {\cal M}_0^{-1} \circ \,\delta{\cal A}_2 [\zeta_i^\pm,\zeta_f^\pm] \rrp_{\bfk, \bfq}  \bigg\} \,,
\end{align}
similarly to what done before. In the formula above, we called
\begin{align}
    \llp {\cal M}_0^{-1} \circ \,\delta{\cal A}_2 [\zeta_i^\pm,\zeta_f^\pm] \rrp_{\bfk, \bfq} &\equiv \int \frac{\dd^3 \bfp}{(2 \pi)^3} \,  {\cal M}^{-1}_0(\bfk, \bfp) \, \delta{\cal A}_2 [\bfp, \bfq, \zeta_i^\pm,\zeta_f^\pm] = \\
    &=\tilde C_{\rm eff}(\bfk)  \delta{\cal A}_2 [-\bfk, \bfq, \zeta_i^\pm,\zeta_f^\pm]\,,
\end{align}
and  the trace operation in Fourier space is defined as 
\begin{align}
    {\rm Tr} \llp A (\bfk, \bfq)\rrp &\equiv \int \frac{\dd^3\bfk\,\dd^3\bfq}{(2\pi)^6}\,(2\pi)^3\delta^{(3)}(\bfk-\bfq)\, A(\bfk,\bfq) \,.
\end{align}

\bigskip
Substituting this expansion into Eq.~\eqref{eq:RhoAvAfterGaussian}, we obtain
\begin{align}\label{eq:RhoAvAfterGaussian2}
    & \rho_{\rm av}(\zeta_f^+,\zeta_f^-;t) = {\cal N} \int \dd \zeta_i^+ \dd \zeta_i^- \: \rho(\zeta_i^+,\zeta_i^-;t_i)\: \times \nonumber \\
    &~~~~~~~~~~~~~~ \times \exp \Bigg\{ i\Omega \llp \delta S_0[\zeta^\pm_i,\zeta^\pm_f] + \alpha^2 \int \frac{\dd^3\bfk}{(2\pi)^3} \, \tilde{C}  (\bfk)  \,\delta{\cal A}_2 [-\bfk,\bfk,\zeta_i^\pm,\zeta_f^\pm]  \rrp \Bigg\} \nonumber\\
    &~~~~~~~~~~~~~~ \times  \exp\Bigg\{ -\frac{\alpha^2 \Omega^2}{2} \int \frac{\dd^3 \bfk }{(2\pi)^3} \delta{\cal A}_1[\bfk,\zeta_i^\pm,\zeta_f^\pm] \tilde{C} (\bfk) \,\delta {\cal A}_1 [-\bfk,\zeta_i^\pm,\zeta_f^\pm]\Bigg\}\,.
\end{align}
For simplicity, and consistently with the perturbative expansion, we retain in the exponents only terms up to order $\alpha^2$. Moreover, the factor $(\det {\cal M}_0)^{-1/2}$ arising from the expansion of $(\det {\cal M})^{-1/2}$ has been absorbed into the overall normalization, since it does not depend on the boundary field configurations on the two Schwinger--Keldysh branches. Truncating the exponentials at this order implies replacing $\tilde C_{\rm eff}\to \tilde C$, effectively dropping  the loop corrections dressing the disorder power spectrum. Note that, by dimensional analysis, $\{ \delta {\cal A}_1, \delta {\cal A}_2 \}$ have dimensions of an action (i.e. $\Omega^{-1}$), while the disorder power spectrum has dimension of a volume, making the exponentials dimensionless.

The structure of Eq.~\eqref{eq:RhoAvAfterGaussian2} makes transparent that the disorder-averaged evolution contains two qualitatively distinct effects, encoded in the two exponentials on the second and third lines. 
Additionally, it shows the non-unitary dynamics is cumulative in propagation time as $T$, as this parameter determines the upper bound on the integrals in ${\cal A}_1, {\cal A}_2$ (see Eqs.~\eqref{eq:A1Fourier} and~\eqref{eq:A2Fourier}). For sufficiently long time the disorder-induced phase fluctuations build up, whereas for short propagation times the same mechanism may remain perturbatively small.

\subsubsection*{First exponential: pure phase factors}
Given the form of ${\cal A}_2$ in Eq.~\eqref{eq:A2Fourier}, and the fact that the disorder power spectrum is real, it is easy to check that 
\be 
\tilde{C}(\bfk) = \llp \tilde{C}(-\bfk) \rrp^* \,, \qquad \delta{\cal A}_2 [\bfk,-\bfk,\zeta_i^\pm,\zeta_f^\pm] = \llp \delta{\cal A}_2 [-\bfk,\bfk,\zeta_i^\pm,\zeta_f^\pm] \rrp^*\,.
\ee 
These two reality conditions imply that the exponential in the second line of Eq.~\eqref{eq:RhoAvAfterGaussian2} is a pure phase. 
It therefore does not produce direct damping or amplification of the density matrix elements, but affects them only through oscillatory cancellations in the integral over the initial configurations. In this sense, it contributes to coherence loss indirectly, by making the averaged evolution increasingly sensitive to relative phase mismatch between the two Schwinger--Keldysh branches. The contribution of $\delta {\cal A}_2$ can be interpreted as a disorder-induced correction to the kernel governing propagation, {\it i.e.} as the analog of a self-energy generated by averaging over the disorder~\cite{Arias:2011pw}. Because the background is static, this correction is analogous to elastic scattering: the wave propagates through a random medium that redistributes phases and amplitudes without allowing energy exchange with the environment. The resulting effect is therefore a modification of coherent propagation through scattering off the inhomogeneities.

\subsubsection*{Second exponential: damping factor}
Similarly, using Eq.~\eqref{eq:A1Fourier}, one finds
\be 
\delta{\cal A}_1 [\bfk,\zeta_i^\pm,\zeta_f^\pm] = \llp \delta{\cal A}_1 [-\bfk, \zeta_i^\pm,\zeta_f^\pm] \rrp^* \,,
\ee 
implying that the combination at the exponent is real. If the disorder power spectrum is positive $\tilde C \geq 0$, then the quadratic form in the third line of Eq.~\eqref{eq:RhoAvAfterGaussian2} is always non-negative. Including the negative multiplicative factor $-\alpha^2\Omega^2/2$, this means that this real exponential term is always damping or equal to 1. 
For $\Omega \neq 0$, then, configurations with non-vanishing branch mismatch are exponentially suppressed, which is the characteristic signature of disorder-induced dephasing, with the suppression enhanced for large values of the parameter.  
The only exception is when the quadratic form vanishes, which corresponds to cases where the boundary field configurations on the two branches coincide, so that $\delta{\cal A}_1=0$. This is the way that decoherence shows up in our formalism: this exponential  is the genuine damping factor generated by averaging the disorder-induced phase fluctuations.

This form also clarifies which scales govern the onset of decoherence. 
From Eq.~\eqref{eq:A1Fourier} one sees that ${\cal A}_1(\bfk)$ is the spatial Fourier transform of a quadratic functional of the free solution $\zeta_0$, and is therefore bilinear in the initial and final boundary data. The difference $\delta{\cal A}_1(\bfk)$ thus measures, for each disorder momentum $\bfk$, how strongly the two Schwinger–Keldysh branches differ in the corresponding Fourier channel. If the correlator $C(\bfx)$ has a finite correlation length $\ell_c$, then its Fourier transform $\tilde C(\bfk)$ is appreciable only over a finite band of momenta, roughly $|\bfk|\lesssim \ell_c^{-1}$. It follows that decoherence is controlled by the overlap between this band and the momentum support of the branch mismatch encoded in $\delta{\cal A}_1$, as understood from the integral of the quadratic form. In particular, the suppression is strongest when the mismatch between the two branches has support on spatial scales comparable to the correlation scale of the disorder, while it is weaker when the relevant Fourier components lie outside the range selected by $\tilde C(\bfk)$.
This result admits a simple physical interpretation. Disorder-induced dephasing is most efficient when the propagating wave resolves the inhomogeneities of the medium. If a single wavelength encompasses many independent fluctuations of the random potential, their effect tends to average out; if instead the wavelength is much shorter than the correlation length, the background is seen as locally smooth. The strongest suppression is therefore expected when the characteristic wavelength of the wave is comparable to the correlation length of the stochastic potential.
Additionally, since ${\cal A}_1$ is built from time derivatives of the free solution, this is also consistent with the interpretation of $\Omega$ as the wave frequency: more rapidly oscillating configurations are more efficiently affected by the random background.

\subsection{Decoherence of Gaussian wave-packets}\label{sec:GaussianPackets}

The decoherence rate is determined by the real exponential appearing in the last line of Eq.~\eqref{eq:RhoAvAfterGaussian2}. In this subsection we study its behavior for Gaussian wave-packets. For sake of simplicity, we restrict our attention to the diagonal elements of the time-evolved averaged density matrix, since these are also the elements needed to compute GW observables (see Eqs.~\eqref{eq:AverageZeta} and~\eqref{eq:AverageZetaZeta}). We consider two nearby initial field configurations, centered around $\bfp_0\pm \Delta \bfp/2$, and final field configurations centered around $\bfp_0$. In order to satisfy the reality condition appropriate for a real field, $(\zeta_{\bfp})^*=\zeta_{-\bfp}$, the momentum-space wave-packets must be even under $\bfp\to -\bfp$. For this reason we choose a superposition of two Gaussians centered at $\pm \bfp_0$, namely
\begin{align}
    \zeta^{f}_\bfp  &= 2\sqrt{2}\left(\frac{\pi}{\sigma^2}\right)^{3/4} \lp  e^{-(\bfp-\bfp_0)^2/(2\sigma^2)} + e^{-(\bfp+\bfp_0)^2/(2\sigma^2)} \rp \,, \label{eq:GaussianF}  \\
    \zeta^{i\pm}_\bfp  &= 2\sqrt{2}\left(\frac{\pi}{\sigma^2}\right)^{3/4} \lp  e^{-(\bfp-\bfp_0\mp \Delta \bfp/2)^2/(2\sigma^2)} + e^{-(\bfp+\bfp_0\pm \Delta \bfp/2)^2/(2\sigma^2)} \rp \,, \label{eq:GaussianIpm}
\end{align} 
with normalization chosen so that
\be
\int \frac{\dd^3 \bfp}{(2\pi)^3}\, |\zeta_\bfp|^2 =1\,.
\ee
Physically, this describes two nearby initial branches in momentum space, each branch being itself a coherent superposition of two opposite momentum lobes. The final state has the same two-lobe structure but is centered on the unshifted value $\pm \bfp_0$. This construction should be contrasted with the one-sided Gaussian packet discussed below for a complex scalar field: in that case a single Gaussian centered at $\bfp_0$ is allowed, since no reality condition relates $\bfp$ and $-\bfp$, and the resulting decoherence pattern is qualitatively different.

We assume that the two initial branches overlap significantly,
\be
|\Delta \bfp|\ll \sigma\,,
\ee
so that the initial state contains non-negligible coherence between them. If instead $|\Delta \bfp|\gg \sigma$, the overlap between the two shifted wave-packets is exponentially suppressed, with a suppression factor of the form $\exp[-|\Delta \bfp|^2/(4\sigma^2)]$. In that regime the two branch states become approximately orthogonal, and the initial density matrix is approximately diagonal in the branch basis spanned by the two packets. Physically, this means that the state already behaves as an incoherent mixture of the two branches before interacting with the disorder, so that there is essentially no branch coherence left for the environment to damp further.

Under this condition one can expand to first order in $\Delta \bfp$ and compute $\delta {\cal A}_1$. Using Eq.~\eqref{eq:DeltaA1Gauss}, one finds
\begin{align}\label{eq:DeltaA1Gaussian}
    &\delta {\cal A}_1 \lp \bfk, \zeta^\pm_i, \zeta_f\rp 
    = \nonumber \\
    &= - 8 \pi^\frac32 \frac{\Delta \bfp }{ \sigma^5}  \cdot \lp 
    (2\bfp_0+ \bfk) e^{-\frac{(\bfp_0 + \bfk/2)^2}{\sigma^2}} +  ( 2\bfp_0 -\bfk) e^{-\frac{(\bfp_0 - \bfk/2)^2}{\sigma^2}} \rp  \int \frac{\dd^3 \bfp}{(2 \pi)^3} \, {\cal A}^{a+b}_1 (\bfp, \bfk)  e^{-\frac{(\bfp + \bfk/2)^2}{\sigma^2}} + \nonumber \\
    &~~~+  16 \pi^\frac32 e^{- \frac{\bfk^2}{4 \sigma^2}} \frac{ \Delta \bfp}{ \sigma^5} \cdot \int \frac{\dd^3 \bfp}{(2 \pi)^3} \, \lp   {\cal A}^{a+b}_1 \lp \bfp + \bfp_0 + \frac{\bfk}{2}, \bfk\rp - {\cal A}^{a+b}_1 \lp \bfp -\bfp_0 -  \frac{\bfk}{2}, \bfk\rp \rp   \bfp  e^{-\frac{\bfp^2}{\sigma^2}} \,,
\end{align}
where we introduced
\be
{\cal A}^{a+b}_1(\bfp,\bfk)\equiv {\cal A}^a_1(\bfp,\bfk)+{\cal A}^b_1(\bfp,\bfk)\,,
\ee
with ${\cal A}^a_1$ and ${\cal A}^b_1$ defined in Eqs.~\eqref{eq:A1A} and~\eqref{eq:A1B}. More details of the derivation are collected in Appendix~\ref{sec:EstimateGaussianPackets}.

To proceed further, we assume that the packets are well localized in momentum space, namely $\sigma\ll 1$. In this limit the second integral is subleading with respect to the first one. Indeed, the Gaussian factor $e^{-\bfp^2/\sigma^2}$ restricts the integral to momenta $|\bfp|\lesssim \sigma$, so that the bracket multiplying $\bfp$ can be expanded around $\bfp=0$. The zeroth-order term gives an odd integrand proportional to $\bfp\, e^{-\bfp^2/\sigma^2}$ and therefore vanishes upon integration. The leading non-vanishing contribution thus starts at first order in the expansion of the kernel difference, so that the integrand is effectively of order $p_i p_j e^{-\bfp^2/\sigma^2}$ and the whole second line scales as $\sigma^3 \nabla_{\bfp}{\cal A}^{a+b}_1$. By contrast, the first line is controlled by the value of the kernel at the center of the packet and scales as $\sigma\,{\cal A}^{a+b}_1$. Hence
\be
\frac{\text{second line}}{\text{first line}}
\sim
\sigma^2 \frac{|\nabla_{\mathbf p}{\cal A}_1^{a+b}|}{|{\cal A}_1^{a+b}|}\,,
\ee 
so that the second line can be neglected provided the kernel does not vary appreciably over momentum scales of order $\sigma$. Physically, this means that we neglect the sensitivity of the disorder kernel to the internal spread of each Gaussian lobe and retain only its value at the center of the packet. In other words, the wave-packet is treated as sufficiently narrow that it probes the disorder only through its central momentum.

Assuming in addition that ${\cal A}^{a+b}_1(\bfp,\bfk)$ is slowly varying over the support of the Gaussian, the remaining integral is dominated by $\bfp\simeq -\bfk/2$, and therefore
\begin{align}
     \delta {\cal A}_1[\bfk,\zeta_i^\pm,\zeta_f]  &\approx - \frac{\Delta \bfp }{ \sigma^2}\cdot \lp 
    (2\bfp_0+ \bfk) e^{-\frac{(\bfp_0 + \bfk/2)^2}{\sigma^2}} +  ( 2\bfp_0 -\bfk) e^{-\frac{(\bfp_0 - \bfk/2)^2}{\sigma^2}} \rp  {\cal A}^{a+b}_1 (-\bfk/2, \bfk)\,.
\end{align}
All the dependence on the propagation time is contained in the factor
\be 
 F_{|\bfk|}(T) \equiv  {\cal A}^{a+b}_1 (-\bfk/2, \bfk) = \frac{|\bfk|}{2 \cos^2\!\big(|\bfk| T / 4\big)} \left( \frac{|\bfk | T}{2} - \sin \frac{|\bfk| T}{2} \right) .
\ee 
This also makes manifest that $\delta {\cal A}_1$ is even under $\bfk\to -\bfk$, as required by its reality.

The decoherence exponent appearing in Eq.~\eqref{eq:RhoAvAfterGaussian2} then becomes
\begin{align}\label{eq:DecoherenceExp}
     -\frac{\alpha^2 \Omega^2}{2 \sigma^4} \int \frac{\dd^3 \bfk }{(2\pi)^3} \tilde{C} (\bfk)  F^2_{|\bfk|}(T) \left[  \Delta \bfp \cdot (2\bfp_0+ \bfk) e^{-\frac{(\bfp_0 + \bfk/2)^2}{\sigma^2}} +  \Delta \bfp \cdot( 2\bfp_0 -\bfk) e^{-\frac{(\bfp_0 - \bfk/2)^2}{\sigma^2}} \right]^2 .
\end{align}

In this form the physical content of the result is transparent. First, the exponent is quadratic in the branch separation $\Delta \bfp$, showing that decoherence suppresses interference between histories that differ in momentum and becomes stronger as the two branches are taken further apart. Second, the square bracket contains two Gaussian contributions, centered respectively around momentum transfers $\bfk\simeq -2\bfp_0$ and $\bfk\simeq 2\bfp_0$. When the square is expanded, one obtains two self-terms and one cross-term. The self-terms describe decoherence associated with each momentum lobe separately, while the cross-term encodes the residual quantum coherence between the two. When this interference term is negligible, the disorder effectively probes the two peaks independently; when it is sizable, the environment remains sensitive to the coherent superposition of the two-lobe state.

This is precisely the point at which the distinction between real and complex fields becomes physically important. For a real field the condition $\zeta^*_{\bfp}=\zeta_{-\bfp}$ forces the state to contain both momentum lobes, and therefore decoherence can receive contributions not only from small momentum transfer but also from transfers able to connect the two peaks, namely $|\bfk|\sim 2|\bfp_0|$. By contrast, for a complex scalar field one may consistently consider a one-sided Gaussian packet centered only at $\bfp_0$. In that case there is no second lobe at $-\bfp_0$, no associated interference term, and for $|\bfp_0|\gg \sigma$ the decoherence exponent is exponentially suppressed because the packet approaches a momentum eigenstate. The real-field packet considered here behaves differently: even when each lobe is sharply localized, decoherence can remain efficient provided the disorder spectrum $\tilde C(\bfk)$ has support at the momentum transfer required to resolve the separation between the two lobes.

Finally, the dependence on the propagation time is entirely encoded in $F_{|\bfk|}(T)$. At short times, when the dominant momentum modes satisfy $|\bfk|T\ll 1$, one finds
\be
F_{|\bfk|}(T)\propto |\bfk|^4 T^3\,,
\ee
and therefore the decoherence exponent scales as $T^6$. Decoherence thus switches on only very slowly at early times. At later times the full trigonometric structure of $F_{|\bfk|}(T)$ must be kept, leading to an oscillatory dependence on $T$ and to possible enhancements for momentum modes close to the zeros of $\cos(|\bfk|T/4)$. Overall, the damping effect is expected to be strongest for larger branch separations $\Delta \bfp$, broader packets in momentum space, longer propagation times, and disorder spectra with appreciable support either at small momentum transfer or at the characteristic transfer $|\bfk|\sim 2|\bfp_0|$ selected by the two-lobe structure required by reality.

A simple parametric estimate can be obtained when the disorder spectrum is isotropic and varies slowly over momentum scales of order $\sigma$. In that case, if $|\bfp_0|\gg \sigma$, the two Gaussian factors in Eq.~\eqref{eq:DecoherenceExp} localize the $\bfk$-integral around the two well-separated regions $\bfk\simeq \pm 2\bfp_0$, each with width of order $\sigma$, and the interference term between the two peaks can be neglected. One may then evaluate the slowly varying functions $\tilde C(\bfk)$ and $F_{|\bfk|}(T)$ at the peak value $|\bfk|\simeq 2|\bfp_0|$, while the remaining Gaussian integral gives only an overall phase-space factor of order $\sigma^3$. Since the vector factor in square brackets is of order $|\Delta \bfp|\,|\bfp_0|$, the decoherence exponential in Eq.~\eqref{eq:DecoherenceExp} takes the form
\begin{align}\label{eq:DecoherenceExpGaussianPackets}
     \approx \exp \llp - \frac{ \alpha^2 \Omega^2  \sigma}{\sqrt{2}\,\pi^{3/2}} \tilde C(2|\bfp_0|)\, F_{2|\bfp_0|}^2(T) \,|\Delta \bfp|^2 \rrp
\end{align}
up to numerical factors of order unity. This estimate makes explicit that the decohering effect is enhanced by increasing the branch separation $|\Delta \bfp|$, the characteristic momentum scale $|\bfp_0|$, the support of the disorder spectrum at momentum transfer $2|\bfp_0|$, and the propagation-time dependence encoded in $F_{2|\bfp_0|}(T)$.

\bigskip
Note that the usual picture of decoherence commonly discussed in open-quantum-system settings differs somewhat from the perspective adopted here. In the standard approach, one typically fixes the same initial state and studies the suppression of the off-diagonal elements of the final density matrix for different final field configurations. By contrast, here we focus on diagonal elements of the final averaged density matrix and analyze how different initial field configurations, corresponding to different histories, are weighted after propagation through the disordered background. The two viewpoints are nevertheless equivalent in the present setup. Indeed, Eq.~\eqref{eq:A1Fourier} shows that $\delta {\cal A}_1$ is symmetric under the exchange of the initial and final field configurations, so the same decoherence mechanism could equally well be described in terms of the damping of off-diagonal final configurations. The choice made here is simply the one most directly suited to the computation of GW observables, since these depend on the diagonal part of the final averaged density matrix, as in Eqs.~\eqref{eq:AverageZeta} and~\eqref{eq:AverageZetaZeta}. In this perspective, decoherence is therefore not represented as the evolution of an initially pure state into a mixed state through the suppression of off-diagonal elements of the final density matrix. Rather, it manifests itself as the dynamical suppression induced by disorder-averaged propagation, of the contributions associated with off-diagonal components of the initial density matrix in the basis of nearby momentum-space wave-packet branches.

\subsection{The influence functional approach}
The same disorder-averaged evolution can be reformulated in the language of the Feynman--Vernon influence functional~\cite{Feynman:1963fq,Calzetta:2008iqa}. This viewpoint is entirely equivalent to the one adopted above, but it makes a complementary aspect of the problem more transparent: averaging over the random background turns the original free theory~\cite{Arias:2011pw} for $\zeta$ into an effective interacting theory on the Schwinger--Keldysh contour. In particular, the effect of the stochastic medium is encoded in a nonlocal self-coupling between the two in-in branches, controlled by the disorder correlator $C(\bfx,\bfy)$.

The starting point is to perform the average over the gravitational potential before carrying out the path integral over $\zeta$. One then obtains
\be
\rho_{\rm av}(\zeta_f^+,\zeta_f^-;t)= \int \dd \zeta_i^+\,\dd \zeta_i^-\,\rho(\zeta_i^+,\zeta_i^-;t_i) \int_{\zeta_i^+}^{\zeta_f^+}\mathcal D\zeta^+ \int_{\zeta_i^-}^{\zeta_f^-}\mathcal D\zeta^-\, e^{i\Omega\left(S_0[\zeta^+]-S_0[\zeta^-]\right)} \,{\cal F}[\zeta^+,\zeta^-]\,,
\ee
where ${\cal F}$ is the influence functional,
\be
{\cal F}[\zeta^+,\zeta^-] \equiv \int \mathcal D\Phi\,\mathcal P[\Phi]\, e^{i\Omega\left(S_{\rm int}[\zeta^+,\Phi]-S_{\rm int}[\zeta^-,\Phi]\right)}
= \exp\!\big(i\Omega S_{\rm inf}[\zeta^+,\zeta^-]\big)\,.
\ee
This defines the influence action $S_{\rm inf}$, which summarizes the entire effect of the stochastic medium on the reduced dynamics after averaging over disorder realizations. In particular, it makes manifest that the role of the environment is to generate couplings between the two Schwinger--Keldysh branches.

For the Gaussian ensemble considered in this work, the functional integral over $\Phi$ can be performed explicitly, as shown in \cite{BOONPAN20121589}. Since the interaction is linear in $\Phi$ and quadratic in $\zeta$, the disorder average generates a quartic and spatially nonlocal term on the closed time contour. One finds, up to equivalent forms related by integrations by parts,
\begin{align}
\label{eq:influencefunc}
&\mathcal{F}[\zeta^+,\zeta^-]= \exp\bigg\{-8\alpha^2\Omega^2\int_0^T \dd^4x\int_0^T \dd^4y\,\times\nonumber\\
&\times
\llp \partial_{t_x}\zeta^+(x)\partial_{t_x}\zeta^+(x) -\, \partial_{t_x}\zeta^-(x)\partial_{t_x}\zeta^-(x) \rrp \, C(\mathbf{x},\mathbf{y}) \llp \partial_{t_y}\zeta^+(y)\partial_{t_y}\zeta^+(y) -\, \partial_{t_y}\zeta^-(y)\partial_{t_y}\zeta^-(y) \rrp  \bigg\}.
\end{align}
This expression makes the physical content of the influence-functional approach particularly transparent. Although for each fixed realization of the background the dynamics remains quadratic, the average over the random potential induces an effective self-interaction for the field on the Schwinger--Keldysh contour. The resulting theory is quartic and nonlocal in space, with coupling strength set by the disorder correlator. In this way, the dephasing effects of the stochastic medium are re-expressed as an effective interaction between the forward and backward branches. This is the quenched-disorder analogue of the influence action familiar from open quantum systems, with the important difference that here the environment is static rather than dynamical.

\section{Conclusions}
In this work, we have applied {\it quenched disorder} techniques to gravitational lensing in order to describe the propagation of a GW through a distribution of weak gravitational lenses. The central idea is to treat the lens distribution as a static random background and to regard the averaged density matrix as the fundamental object from which observables are computed. This framework goes beyond the geometric-optics limit and extends the usual wave-optics treatment from a single lens to the more realistic case of a random lens distribution. More generally, it provides a natural starting point for studying how stochastic lensing environments affect GW propagation and may be relevant for the description of GW-background anisotropies.

Within this approach, we constructed a path-integral representation of the disorder-averaged density matrix using the Schwinger–Keldysh formalism, which consistently accounts for forward and backward time evolution while naturally implementing the ensemble average over lens realizations. The resulting description treats interference, diffraction, and statistical fluctuations of the medium within a single framework.

Our main result is that the averaged density matrix at the final time decomposes naturally into two contributions: a damping term and a pure phase. The first is an exponential quadratic form built from $\delta{\cal A}_1$ and the effective kernel ${\cal M}^{-1}$, which governs the suppression of the density-matrix elements and encodes the coherence loss induced by disorder fluctuations. The second is a purely oscillatory contribution, determined primarily by the mismatch in free propagation between the two Schwinger–Keldysh branches and corrected by a term associated with $\delta{\cal A}_2$. This correction can be interpreted as a disorder-induced modification of the propagation kernel, analogous to a self-energy. Because the background is static, it corresponds to elastic scattering and modifies coherent propagation without energy exchange.
The dependence on the parameter $\Omega$ further clarifies the physical picture. In the small-$\Omega$ regime, both the damping and the oscillatory phase are weak, so propagation approaches the free case. In the opposite limit, $\Omega \to \infty$, rapidly oscillating phases select stationary configurations, while the quadratic exponential strongly suppresses the off-diagonal elements of the averaged density matrix.
Our results also identify the scales controlling the onset of coherence loss. The quantity $\delta{\cal A}_1(\mathbf{k})$ measures the branch mismatch in each Fourier channel, while the disorder correlator $\tilde C(\mathbf{k})$ selects the momenta that contribute to the averaging. The suppression is therefore controlled by the overlap between these two quantities. In the explicit examples considered here, the effect is strongest when the wavelength of the propagating wave is comparable to the correlation length of the disorder, and weaker when the two scales are widely separated. We illustrated these features explicitly for Gaussian wave packets.

A natural next step is to extend the formalism to more general interaction structures, to non-Gaussian disorder statistics, and to investigate more systematically its potential as a probe of the matter power spectrum in different physical scenarios. Although our discussion has been framed in terms of GW lensing, the derivation is more general and applies to any system described by the class of actions from which we started, making the framework potentially relevant to a broader set of wave-propagation problems in disordered backgrounds.

\acknowledgments
We would like to thank Luca Dell’Anna for helpful discussions, and the Gran Sasso Science Institute (GSSI) for hosting the workshop ``The Complexity of the Cosmos’’, during which this work was conceived.
A.G. is supported by funds provided by the Center for Particle
Cosmology at the University of Pennsylvania.
RA, NB and SM acknowledge partial financial support by the MUR Departments of Excellence grant “Quantum Frontiers” of the Physics and Astronomy Department of Padova University. Additionally, NB and SM acknowledge partial financial support by the COSMOS network (www.cosmosnet.it) through the ASI (Italian Space Agency) Grants 2016-24-H.0, 2016-24-H.1-2018 and 2020-9-HH.0.

\appendix
\section{Evaluation of the classical action}
\label{App:A}

This appendix collects the technical steps required to evaluate the classical action $S_{\rm cl}$ perturbatively in the coupling $\alpha$. This is given by the three contributions in Eq.~\eqref{actionclassical}, which we tackle one at the time in Secs.~\ref{AppA:free},~\ref{AppA:first} and~\ref{AppA:second} respectively.

\subsection{Free classical solution and action}
\label{AppA:free}
In this Section we provide the action evaluated on the classical trajectory evaluated at lowest order in $\alpha$, i.e we give explicitly $\zeta_0$ in Eq.~\eqref{Eq:perturbativeexpansionzeta} and $S_0$ in Eq.~\eqref{actionclassical}. 
It is easy to recognize that the solution for $\zeta_0$ with the boundary conditions in Eq.~\eqref{eq:BCzeta0} is 
\be\label{eq:zeta0_solution}
\zeta_0(\bfx,t)= \int\frac{\dd^3 \bfk}{(2\pi)^3} e^{i\bfk\cdot\bfx} \llp \frac{\zeta^i_{\bfk}\sin\!\big(|\bfk|(T-t)\big)}{\sin(|\bfk| T)} + \frac{\zeta^f_\bfk\sin(|\bfk| t)}{\sin(|\bfk| T)} \rrp\,,
\ee 
where 
\be 
\zeta^{i,f}_\bfk = \int \dd^3 \bfx \,, e^{- i \bfk \cdot \bfx} \zeta_{i,f}(\bfx)\,,
\ee 
are the Fourier transform of the boundary field configurations.
Using Eq.~\eqref{eq:action in spacetime}, and keeping terms at order $\alpha^0$ yields 
\be \label{eq:S0classical}
S_0[\zeta_i, \zeta_f] =  \int \frac{\dd^3 \bfk}{(2\pi)^3} \frac{|\bfk|^2 T}{\sin \lp |\bfk| T\rp} \llp \zeta^i_\bfk \zeta^i_{-\bfk}  + \zeta^f_\bfk \zeta^f_{-\bfk} - \cos (|\bfk| T) \lp \zeta^i_\bfk \zeta^f_{-\bfk}  + \zeta^f_\bfk \zeta^i_{-\bfk} \rp  \rrp \,.
\ee 
Note that the reality of $\zeta$ imposes that $(\zeta^{i,f}_\bfk)^* = \zeta^{i,f}_{-\bfk}$; hence, the expression above is real.

\subsection{First-order correction} \label{AppA:first}

In this Section we provide the action evaluated on the classical trajectory evaluated at first order in $\alpha$, i.e we give explicitly $\zeta_1$ in Eq.~\eqref{Eq:perturbativeexpansionzeta}, with Dirichlet boundary condition, and $S_1$ in Eq.~\eqref{actionclassical}. 
Using Eq.~\eqref{eq:action in spacetime}, and keeping terms at order $\alpha^1$ yields 
\begin{align}\label{eq:S1classical}
    S_1 [\zeta_i, \zeta_f, \Phi] &= \int^T_0 \dd^4 x \llp 2 \partial^\mu \zeta_0 \partial_\mu \zeta_1 + 4 \Phi (\bfx) (\partial_t \zeta_0 )^2\rrp \,.
\end{align}
Going to Fourier space, using Eq.~\eqref{eq:zeta1FromGreen} with $\zeta_0$ expressed as in Eq.~\eqref{eq:zeta0_solution} and the free propagator in Eq.~\eqref{eq:FreeGFkt} (or its resummed version in ~\eqref{Eq:free_prop_resummed}), we can write this as 
\be \label{eq:S1classicalFourier}
    S_1 [\zeta_i, \zeta_f, \Phi] = \int \frac{\dd^3 \bfk}{(2 \pi)^3} \: \tilde{\Phi}(\bfk) \, \times {\cal A}_1 \lp \bfk, \zeta_i, \zeta_f\rp
\ee 
with 
\be \label{eq:A1FourierTemp}
{\cal A}_1 \lp \bfk, \zeta_i, \zeta_f\rp \equiv 4 \int^T_0 \dd^4 x \, e^{i \bfk \cdot \bfx} \llp (\partial_{t_x} \zeta_0(x) )^2 - 2 \partial^2_{t_x} \zeta_0 (x) \int^T_0 \dd^4 x'  \, \partial'^{\mu} \zeta_0 (x') \, \partial'_{\mu} \Delta_0 (x',x)  \rrp
\ee 
where $\partial'_{\mu}$ is the partial derivative with respect to $x'$. From this expression one can easily obtain the explicit expression in terms of the values of the boundary field configurations. A straightforward computation, using Eqs.~\eqref{eq:zeta0_solution} and~\eqref{Eq:free_prop_resummed} shows that the second term in Eq.~\eqref{eq:A1FourierTemp} actually vanishes, leaving
\be
{\cal A}_1 \lp \bfk, \zeta_i, \zeta_f\rp \equiv 4 \int^T_0 \dd^4 x \, e^{i \bfk \cdot \bfx}  (\partial_{t_x} \zeta_0(x) )^2 \,,
\ee 
and thus 
\begin{align}\label{eq:A1Fourier}
    {\cal A}_1 \lp \bfk, \zeta_i, \zeta_f\rp 
    &=  \int \frac{\dd^3 \bfp}{(2 \pi)^3} \, \Bigg\{ \llp \zeta^i_\bfp \zeta^i_{-\bfp - \bfk} + \zeta^f_\bfp \zeta^f_{-\bfp - \bfk} \rrp {\cal A}^a_1 (\bfp, \bfk) +  \llp \zeta^i_\bfp \zeta^f_{-\bfp - \bfk} + \zeta^f_\bfp \zeta^i_{-\bfp - \bfk} \rrp {\cal A}^b_1 (\bfp, \bfk)  \Bigg\}
\end{align}
where we have defined
\begin{align}
    {\cal A}^a_1 (\bfp, \bfk) &\equiv \frac{4 \, |\bfp||\bfp + \bfk|}{|\bfp|^2 - |\bfp + \bfk|^2 } \Big( |\bfp| \cot \llp |\bfp + \bfk| T  \rrp - |\bfp+ \bfk| \cot \llp |\bfp | T  \rrp\Big) \,, \label{eq:A1A}\\
    {\cal A}^b_1 (\bfp, \bfk) &\equiv \frac{4 \, |\bfp||\bfp + \bfk|}{|\bfp|^2 - |\bfp + \bfk|^2 }\lp  \frac{|\bfp+ \bfk|}{\sin[|\bfp| T]} - \frac{|\bfp|}{\sin[|\bfp+ \bfk| T]}\rp\,. \label{eq:A1B}
\end{align}

\subsection{Second-order correction}
\label{AppA:second}

In this Section we provide the action evaluated on the classical trajectory evaluated at second order in $\alpha$, i.e we give explicitly  $S_2$ in Eq.~\eqref{actionclassical}. 
Using Eq.~\eqref{eq:action in spacetime}, and keeping terms at order $\alpha^2$ yields 
\begin{align}\label{eq:S2classical}
    S_2 [\zeta_i, \zeta_f, \Phi] &= \int^T_0 \dd^4 x \llp 2 \partial^\mu \zeta_0 \partial_\mu \zeta_2 + \partial^\mu \zeta_1 \partial_\mu \zeta_1 + 8 \Phi (\bfx) \partial_t \zeta_0 \partial_t \zeta_1 \rrp \,.
\end{align}
Going to Fourier space, using Eqs.~\eqref{eq:zeta1FromGreen} and~\eqref{eq:zeta2FromGreen}, with $\zeta_0$ expressed as in Eq.~\eqref{eq:zeta0_solution} and the free propagator in Eq.~\eqref{eq:FreeGFkt} (or its resummed version in ~\eqref{Eq:free_prop_resummed}), we can write this as
\be \label{eq:S2classicalFourier}
    S_2 [\zeta_i, \zeta_f, \Phi] = \int \frac{\dd^3 \bfk \dd^3 \bfq} {(2 \pi)^6} \:\: \tilde{\Phi}(\bfk) \tilde{\Phi}(\bfq) \, \times {\cal A}_2 \lp \bfk, \bfq, \zeta_i, \zeta_f\rp \,,
\ee 
with 
\begin{align} 
{\cal A}_2 \lp \bfk, \bfq, \zeta_i, \zeta_f\rp &\equiv 16 \int^T_0 \dd^4 x \dd^4 y\, e^{i \bfk \cdot \bfx + i \bfq \cdot \bfy} \Bigg[2 \int^T_0 \dd^4 x' {\partial'}^\mu \zeta_0 (x') \partial'_\mu \Delta_0(x',x) \partial^2_{t_x} \Delta_0(x,y)\partial^2_{t_y} \zeta_0(y) + \nonumber \\
&~~~+ \int^T_0 \dd^4 x' {\partial'}^\mu \Delta_0 (x', x) \, \partial'_\mu \Delta_0 (x', y) \, \partial_{t_x}^2 \zeta_0 (x)  \partial_{t_y}^2 \zeta_0 (y)   + \nonumber \\
&~~~-2 \partial_{t_y} \zeta_0 (y) \,  \partial_{t_y} \Delta_0(y,x) \, \partial^2_{t_x} \zeta_0(x) \Bigg]\,.
\end{align}
Similarly to the first order correction, many of these integrals can be simplified using the definitions of $\Delta_0$ and $\zeta_0$. Indeed, a straightforward manipulations yields
\be \label{eq:A2Fourier}
    {\cal A}_{2} \lp \bfk, \bfq, \zeta_i, \zeta_f\rp  = 16 \int^T_0 \dd^4 x \dd^4 y e^{i \bfk \cdot \bfx + i \bfq \cdot \bfy} \llp \Delta_0 (y,x) \partial^2_{t_x} \zeta_0 (x)\partial^2_{t_y} \zeta_0 (y)\rrp\,.
\ee


\section{Evaluation of the Loop-like corrections}
\label{App:B}
In this section, we show the explicit calculations that have been performed to obtain Eq.~\eqref{eq:dressed_TEO_2order}.

We start by computing explicitly the tadpole contribution
\begin{align}
   -\frac{1}{2}\text{Tr}\llp 4\alpha \Phi({\bfx})\partial_t^2\Delta_0(x,y)\rrp=&2\alpha \Bigg\{\int \frac{d^3x d^3k'}{(2\pi)^3}e^{i{\bf k}'{\bfx}}\Tilde\Phi({\bf k}')\int \frac{d^3y d^3k}{(2\pi)^3}e^{i{\bfk}({\bfx}-{\bfy})}\delta^{(3)}({\bfx}-{\bfy})\times\nonumber\\ &\times\sum_{n=1}^{\infty}\omega_n^2\int_0^Tdt dt' \delta(t-t')\frac{2}{T}\frac{\sin(\omega_n t)\sin(\omega_n t')}{\omega_n^2-|\bfk|^2}\Bigg\}\,.
\end{align}
Some of the contributions in the equation above simplify. In particular we have that:
\begin{align}
    &\int d^3x e^{i{\bfk}'{\bfx}}=(2\pi)^3 \delta^{(3)}({\bfk}')\,,\\
    &\int_0^T dt' \delta(t-t')\sin(\omega_n t')= \begin{cases}
                                                    \sin(\omega_n t)\qquad \text{if} \quad t\in (0,T)\,\\
                                                    0 \qquad \text{otherwise}\,.
                                                    \end{cases}
\end{align}
Since, by assumption, $t\in(0,T)$, we are left with
\begin{align}
\label{Eq:tadpole}
   &-\frac{1}{2}\text{Tr}\llp 4\alpha \Phi({\bfx})\partial_t^2\Delta_0(x,y)\rrp=2\alpha \tilde\Phi(0)\int \frac{d^3k}{(2\pi)^3}\sum_{n=1}^{\infty}\frac{\omega_n^2}{ \omega_n^2-|\bfk|^2}\int_0^T dt \frac{2}{T}\sin^2(\omega_n t)\nonumber\\
   &=2\alpha \tilde\Phi(\bfk=0)\int \frac{d^3k}{(2\pi)^3}\sum_{n=1}^{\infty}\frac{\omega_n^2}{\omega_n^2-|\bfk|^2}\,
\end{align}
where in the last step we used the orthonormality condition. \\
Eq.~\eqref{Eq:tadpole} correspond to the tadpole correction to the propagator. The fact that we have $\tilde\Phi(0)$ entering means that the tadpole loop is sensitive only to the constant part of the background, {\it i.e.} the average value of $\Phi$ over space.

Analogously, one can compute the second contribution of equation \eqref{Eq:Loop corrections}:
\begin{align}
    &\frac{1}{2}8\alpha^2\text{Tr}\left[ \int_0^T d^4 z\, \Phi(\bfx)\, \partial_t^2 \Delta_0(x, z)\, \Phi(\textbf{z})\, \partial_{t''}^2 \Delta_0(z, y)\right]=\nonumber\\
    =&4\alpha^2\int \frac{d^3q\,d^3k}{(2\pi)^6}\tilde\Phi(\textbf{q})\tilde\Phi(-\textbf{q})\sum_{n=1}^{+\infty}\frac{\omega_n^4 }{(\omega_n^2-|\bfk|^2)(\omega_n^2-|({\bfk}+\textbf{q})|^2)}\,,
    \end{align}
    where we used the orthonormality condition and we observed that
\begin{align}
    \begin{cases}
    &\int_0^Tdt\,\sin(\omega_n t)\sin(\omega_m t)=\frac{T}{2}\delta_{nm}\,\\
        &\int d^3x\,\exp\left\{i{\bfx}\cdot({\textbf{q}}+{\bfk}-{\bfk}')\right\}=(2\pi)^3\delta^{(3)}({\bfk}'-({\textbf{q}}+{\bfk}))\,,\\
        &\int d^3z\,\exp\left\{i{\textbf{z}}\cdot({\textbf{q}'}-{\bfk}+{\bfk}')\right\}=(2\pi)^3\delta^{(3)}({\textbf{q}'}-({\bfk}-{\bfk}'))\,.
    \end{cases}
\end{align}

\section{Gaussian wave-packets}\label{sec:EstimateGaussianPackets}
In this section we provide the computations backing Section~\ref{sec:GaussianPackets} where we compute the decoherence exponential
\be 
\exp\Bigg\{ -\frac{\alpha^2 \Omega^2}{2} \int \frac{\dd^3 \bfk }{(2\pi)^3} \delta{\cal A}_1[\bfk,\zeta_i^\pm,\zeta_f] \tilde{C} (\bfk) \,\delta {\cal A}_1 [-\bfk,\zeta_i^\pm,\zeta_f^\pm]\Bigg\} \,,
\ee 
on the diagonal elements of the field basis.
The form of $\delta {\cal A}_1$ can be found from Eqs.~\eqref{eq:A1Fourier} by taking the difference between the $(+)$ and $(-)$ branches. We find 
\begin{align}\label{eq:DeltaA1Gauss}
    \delta {\cal A}_1 \lp \bfk, \zeta^\pm_i, \zeta_f\rp 
    &=  \int \frac{\dd^3 \bfp}{(2 \pi)^3} \, \Bigg\{ \llp \zeta^{i+}_\bfp \zeta^{i+}_{-\bfp - \bfk} - \zeta^{i-}_\bfp \zeta^{i-}_{-\bfp - \bfk} \rrp {\cal A}^a_1 (\bfp, \bfk)  + \nonumber \\
    &~~~~~~~~~ +  \llp \zeta^{f}_{-\bfp - \bfk} \lp \zeta^{i+}_\bfp - \zeta^{i-}_\bfp  \rp + \zeta^{f}_\bfp \lp \zeta^{i+}_{-\bfp - \bfk} -  \zeta^{i-}_{-\bfp - \bfk} \rp  \rrp {\cal A}^b_1 (\bfp, \bfk)  \Bigg\}
\end{align}
with ${\cal A}^a_1$ and ${\cal A}^b_1$ given in Eqs.~\eqref{eq:A1A},~\eqref{eq:A1B}.
Taking Gaussian wave-packets as in Eqs.~\eqref{eq:GaussianF} and~\eqref{eq:GaussianIpm}, we expand the initial field configurations for $|\Delta \bfp| \ll \sigma$ as 
\be 
\zeta^{i\pm}_\bfp \approx \zeta^{f}_\bfp \pm \Delta \bfp \cdot \bfu_\bfp \,,
\ee 
with 
\be 
\bfu_\bfp = 2\sqrt{2}\left(\frac{\pi}{\sigma^2}\right)^{3/4} \frac{1}{2 \sigma^2}\llp (\bfp - \bfp_0) e^{ \lp -\frac{(\bfp - \bfp_0)^2}{2\sigma^2} \rp} -(\bfp + \bfp_0)  e^{ \lp -\frac{(\bfp + \bfp_0)^2}{2\sigma^2} \rp } \rrp\,.
\ee 
Then, it is easy to realize that the combinations in square brakets in $\delta {\cal A}_1$ are the same at linear order in $|\Delta \bfp|$
\begin{align}
    \zeta^{i+}_\bfp \zeta^{i+}_{-\bfp - \bfk} - \zeta^{i-}_\bfp \zeta^{i-}_{-\bfp - \bfk}   &\approx  2 \Delta \bfp \cdot  \lp \zeta^{f}_\bfp \, \bfu_{-\bfp -\bfk} +  \zeta^{f}_{-\bfp -\bfk} \,  \bfu_\bfp \rp \\
    \zeta^{f}_{-\bfp - \bfk} \lp \zeta^{i+}_\bfp - \zeta^{i-}_\bfp  \rp + \zeta^{f}_\bfp \lp \zeta^{i+}_{-\bfp - \bfk} -  \zeta^{i-}_{-\bfp - \bfk} \rp &\approx  2 \Delta \bfp \cdot  \lp \zeta^{f}_\bfp \, \bfu_{-\bfp -\bfk} +  \zeta^{f}_{-\bfp -\bfk} \,  \bfu_\bfp \rp \,,
\end{align}
and in particular 
\begin{align}
    2 \Delta &\bfp \cdot  \lp \zeta^{f}_\bfp \, \bfu_{-\bfp -\bfk} +  \zeta^{f}_{-\bfp -\bfk} \,  \bfu_\bfp \rp = \nonumber \\
    &= - \frac{8}{ \sigma^2} \left(\frac{\pi}{\sigma^2}\right)^{\frac32}  \Delta \bfp \cdot \Bigg\{  
    (2\bfp_0+ \bfk) e^{-\frac{(\bfp + \bfk/2)^2}{\sigma^2}} e^{-\frac{(\bfp_0 + \bfk/2)^2}{\sigma^2}} +  ( 2\bfp_0 -\bfk) e^{-\frac{(\bfp + \bfk/2)^2}{\sigma^2}} e^{-\frac{(\bfp_0 - \bfk/2)^2}{\sigma^2}}  + \nonumber \\
    &~~~~~+  (2\bfp + 2\bfp_0+\bfk) e^{-\frac{(\bfp + \bfp_0 + \bfk/2)^2}{\sigma^2}} e^{- \frac{\bfk^2}{4 \sigma^2}} + ( 2\bfp_0 - 2\bfp -\bfk) e^{-\frac{(\bfp - \bfp_0 + \bfk/2)^2}{\sigma^2}} e^{- \frac{\bfk^2}{4 \sigma^2}}  \Bigg\}\,.
\end{align}
Plugging back into the integral, one obtains Eq.~\eqref{eq:DeltaA1Gaussian}.

\section{Power spectrum}
We want to obtain an explicit prediction for the disordered-averaged power spectrum. We work by expanding the \emph{normalized} expectation value perturbatively in the disorder strength. Let us define the final momentum-space power spectrum as
\be
P^\zeta_{\rm av} (\bfk, \bfp,t_f) \equiv  \frac{\langle \zeta^f_{\bfk}\zeta^f_{\bfp} \rangle_{\rm av}}{\langle 1\rangle_{\rm av}} =
\frac{\int \dd\zeta_i^+\dd\zeta_i^- \dd\zeta_f \; \rho(\zeta_i^+,\zeta_i^-;t_i)\, e^{\,i\Omega \delta S_0[\zeta_i^\pm,\zeta_f]}\, e^{S_{\rm dis}[\zeta_i^\pm,\zeta_f]}\, \zeta^f_{\bfk}\zeta^f_{\bfp}} {\int \dd\zeta_i^+\dd\zeta_i^- \dd\zeta_f \; \rho(\zeta_i^+,\zeta_i^-;t_i)\, e^{\,i\Omega \delta S_0[\zeta_i^\pm,\zeta_f]}\, e^{S_{\rm dis}[\zeta_i^\pm,\zeta_f]}}\,,
\ee
where 
\be 
S_{\rm dis} [\zeta_i^\pm,\zeta_f^\pm] \equiv \alpha^2 \int \frac{\dd^3\bfk}{(2\pi)^3} \, \tilde{C}  (\bfk)  \Bigg\{ i \Omega  \delta{\cal A}_2 [-\bfk,\bfk,\zeta_i^\pm,\zeta_f^\pm]   -\frac{\Omega^2 }{2} \delta{\cal A}_1[\bfk,\zeta_i^\pm,\zeta_f^\pm] \,\delta {\cal A}_1 [-\bfk,\zeta_i^\pm,\zeta_f^\pm] \Bigg\}
\ee 
as understood from Eq.~\eqref{eq:RhoAvAfterGaussian2}. Let us also introduce the power spectrum in the absence of the gravitational potential 
\be 
P^\zeta_{0} (\bfk, \bfp,t_f) \equiv \frac{\langle \zeta^f_{\bfk}\zeta^f_{\bfp} \rangle_{0}}{\langle 1\rangle_{0}} = \frac{\int \dd\zeta_i^+\dd\zeta_i^- \dd\zeta_f \; \rho(\zeta_i^+,\zeta_i^-;t_i)\, e^{\,i\Omega \delta S_0[\zeta_i^\pm,\zeta_f]}\, \zeta^f_{\bfk}\zeta^f_{\bfp}} {\int \dd\zeta_i^+\dd\zeta_i^- \dd\zeta_f \; \rho(\zeta_i^+,\zeta_i^-;t_i)\, e^{\,i\Omega \delta S_0[\zeta_i^\pm,\zeta_f]}\,}\,,
\ee 
where $\langle \dots \rangle_{0}$ is the average computed keeping the density matrix at $\alpha^0$ order. Assuming a translationally invariant and isotropic initial density matrix, the free measure is translationally invariant, leading to the usual 
\be 
P^\zeta_{0} (\bfk, \bfp,t_f)  = (2 \pi)^3 \delta^{(3)} (\bfk + \bfp) P^\zeta_{0} (|\bfk|,t_f)\,.
\ee 
Expanding the disorder exponential as $e^{S_{\rm dis}} \approx 1 + S_{\rm dis} + {\cal O}(\alpha^4)$, we find 
\be \label{eq:PzetaConnected}
P^\zeta_{\rm av} (\bfk, \bfp,t_f)  =(2 \pi)^3 \delta^{(3)} (\bfk + \bfp) P^\zeta_{0} (|\bfk|,t_f)  + \llp \frac{\langle \zeta^f_{\bfk}\zeta^f_{\bfp} S_{\rm dis}\rangle_{\rm 0} }{\langle 1\rangle_{0}}-  \frac{\langle \zeta^f_{\bfk}\zeta^f_{\bfp} \rangle_{\rm 0} }{\langle 1\rangle_{0}}\frac{\langle  S_{\rm dis}\rangle_{\rm 0} }{\langle 1\rangle_{0}}\rrp + {\cal O}(\alpha^4)\,.
\ee 
where in the square bracket we recognize the connected  correlator, which can be computed explicitly given $S_{\rm dis}$ and a choice of the initial density matrix and disorder power spectrum. 
Note that, because  $S_{\rm dis}$ contains terms without $\zeta^f_\bfk$ and linear and quadratic in $\zeta^f_\bfk$, the connected correlators that have to be computed are higher point functions of the form $\langle \zeta^f_\bfk \zeta^f_\bfp \dots \zeta^f_\bfq \rangle_{0}$.
Since the measure at $\alpha^0$ order is translationally invariant, every free contraction is diagonal in momentum space, and therefore each contribution is proportional to $(2\pi)^3\delta^{(3)}(\bfk+\bfp)$. Hence, the disorder-averaged power spectrum retains the diagonal form
\be\label{eq:Pzetadiagonal}
P^\zeta_{\rm av} (\bfk, \bfp,t_f) = (2 \pi)^3 \delta^{(3)} (\bfk + \bfp) \llp  P^\zeta_{0} (|\bfk|,t_f) +\delta P^\zeta (|\bfk|,t_f) \rrp +{\cal O}(\alpha^4)\,,
\ee
with $\delta P^\zeta$ obtained by computing the correlators in Eq.~\eqref{eq:PzetaConnected} explicitly.

Computing the averages in  Eq.~\eqref{eq:PzetaConnected} in full generality is highly non-trivial. It would be interesting however to do so and compare the result of the power spectrum obtained from the Born approximation of the GW. Indeed, one could compute as 
\begin{align}
    P^\zeta_{\rm av} (\bfk, \bfp,t_f) &= \langle (\zeta_{0, \bfk} + \alpha \zeta_{1, \bfk} + \alpha^2 \zeta_{2, \bfk} )  (\zeta_{0, \bfp} + \alpha \zeta_{1, \bfp} + \alpha^2 \zeta_{2, \bfp} )\rangle_0 = \nonumber \\
    &=(2 \pi)^3 \delta^{(3)} (\bfk + \bfp)   P^\zeta_{0} (|\bfk|,t_f) + \alpha^2 \llp \langle   \zeta_{1, \bfk} \zeta_{1, \bfp} \rangle_0 +  \langle   \zeta_{0, \bfk} \zeta_{2, \bfp} \rangle_0 +  \langle   \zeta_{2, \bfk} \zeta_{0, \bfp} \rangle_0 \rrp\,.
\end{align}
where $\zeta_i$ are the ones defined in Eq.~\eqref{Eq:perturbativeexpansionzeta}.
This is the same perturbative strategy commonly adopted in the literature on scalar-induced gravitational waves~\cite{Kugarajh:2025pjl,Domenech:2021ztg,Matarrese:1993zf,Matarrese:1997ay,Domenech:2020kqm,Baumann:2007zm,Ananda:2006af} and scalar-tensor induced gravitational waves~\cite{Picard:2023sbz,Bari:2021xvf,Bari:2023rcw,Garoffolo:2022usx}, where one expands the wave field order by order and then evaluates the corresponding two-point function. In our framework, however, the full propagation between the initial and final states is computed exactly first, and the expansion is performed only afterward at the level of the disorder-averaged observable. For this reason, the two constructions need not agree at finite perturbative order, since the present approach effectively retains cumulative propagation effects that are not manifest in a truncated Born expansion. It would be interesting to compare the two results

\bibliography{bibliography}
\bibliographystyle{JHEP}
\end{document}